\documentclass[aps,pra, twocolumn,showpacs]{revtex4}
\newcommand{\la}{\lambda}

\newcommand{\cn}{{\mathrm{cn}} }
\newcommand{\dn}{{\mathrm{dn}} }
\newcommand{\sn}{{\mathrm{sn}} }

\newcommand{\tm}{\tilde{m}}
\newcommand{\tkappa}{\tilde{\kappa}}

\newcommand{\tx}{\tilde{x}}
\newcommand{\tr}{\tilde{r}_\bot}

\newcommand{\tsigma}{\tilde{\sigma}}
 \newcommand{\tmp}{\tilde{m}^\prime}

\usepackage{epsfig}
\usepackage{graphicx}
\usepackage{amsmath}
\usepackage{amssymb}
\usepackage{dcolumn}

\begin{document}
%\twocolumn[\hsize\textwidth\columnwidth\hsize
%\csname@twocolumnfalse%
%\endcsname
\draft
\title {Management of matter waves in optical lattices by means of Feshbach resonance}
%{Feshbach resonance management of matter waves in optical lattices}
\author{V.A. Brazhnyi$^1$}
\email{brazhnyi@cii.fc.ul.pt}
\author{V.V. Konotop$^{1,2}$}
\email{konotop@cii.fc.ul.pt}
%\address{
\affiliation{$^1$Centro de F\'{\i}sica Te\'orica e Computacional,
Universidade de Lisboa,
 Complexo Interdisciplinar, Av. Prof. Gama Pinto 2, Lisboa 1649-003,
 Portugal \\
 $^2$ Departamento de F\'{\i}sica, Faculdade de Ci\^encias,
Universidade de Lisboa, Campo Grande, Ed. C8, Piso 6, Lisboa
1749-016, Portugal.
 }
%\date{\today}


\begin{abstract}
Mean-field dynamics of Bose-Einstein condensates loaded in an optical lattice, confined by a parabolic potential, and subjected to change of a scattering length by means of the Feshbach resonance is considered.
The system is described by the Gross-Pitaevskii (GP) equation with varying nonlinearity, which in a number of cases is reduced to one-dimensional (1D) perturbed nonlinear Schr\"{o}dinger (NLS) equations, particular form of which depends on relation among parameters of the problem. 
We analytically describe adiabatic dynamics of periodic solutions of the respective NLS equations, provide numerical study of 1D models confined by a parabolic trap, and carry out numerical simulations of the matter wave dynamics within the framework of the radially symmetric 3D GP equation. Special attention is paid to processes of generation of trains of bright and dark matter solitons from initially periodic waves.
The results of the 1D approximation are compared with direct numerical simulation of the original 3D GP eqation, showing remarkable coincidence for definite regions of parameters.
\end{abstract}

\maketitle

\section{Introduction}

During the last decade Bose-Einstein condensates (BECs) attracted a great deal of attention of the physical community (see e.g. review papers \cite{reviews,MO,BK_rev} and references therein). 
Creation and manipulation of spatially localized and periodic matter waves have become the issues of particular importance, optical lattices (OLs) being viewed as one of the most promising tools for management \cite{MO,BK_rev,BO,bright}. 
On the other hand, it has been predicted theoretically \cite{FR_theory} and confirmed experimentally \cite{bright,FR} that another tool -- the Feshbach resonance (FR) -- can be effectively employed for manipulating BECs. 
It also has been suggested that by changing external magnetic field (i.e. by controlling the magnitude and even sign of the s-wave scattering length) one can generate trains of solitons starting with a periodic matter wave \cite{prl90AKKB}, create stable localized structures such as breathers, 2-soliton states or nearly static states with nested dark solitons \cite{prl90KTFM}, compress solitary waves up to high densities \cite{AS}, generate shock waves \cite{shock} in BECs and create single bright and dark solitons starting with the linear oscillator eigenstates \cite{KKVF}. 
Subsequently, a natural question about combined effect of FR and periodic external potential can be posed. 
This issue has already been addressed in Ref.\cite{ATMK}, where effect of variations of the scattering length on dynamics of a soliton of a discrete nonlinear Schr\"{o}dinger (NLS) equation, which simulates an array of BECs in a deep OL within the framework of the tight-binding approximation~\cite{ABDKS}, was considered. 
However, as it was shown in \cite{AKKS}, the latter approximation does not reproduce all features of dynamics of the underlining continuous system, what requires extending consideration beyond the tight-binding limit.

The aim of the present paper is to consider combined effect of FR and of an OL on dynamics of a BEC within the framework of reduced one-dimensional (1D) and governed by truly 3D mean-filed models.
More specifically, we extend the ideas reported in Ref.~\cite{prl90AKKB} to cases of BECs embedded in OLs and consider the effect of FR on dynamics of nonlinear Bloch states resulting in creation of trains of bright and dark solitons. 

The problem allows rather complete analytical description in a 1D case. 
That is why we start by listing in Sec.~\ref{1Dnls} the main 1D reductions of the 3D Gross-Pitaevskii (GP) equation for different  relations among characteristic scales of the problem. 
Next, in Sec.~\ref{NLS} we recall some relevant facts of the adiabatic theory of the NLS equation 
providing some details to the earlier results of Ref.~\cite{prl90AKKB}.
This will lead us to simple picture of the effect of varying nonlinearity on 1D periodic solutions when the lattice constant is small enough. 
In Sec.~\ref{ini} adiabatic approximation is applied to describe dynamics of periodic mater waves loaded in a periodic potential and affected by FR.
In Sec.~\ref{num1D} we provide numerical study of 1D dynamics of solutions describing BEC in an OL where boundary effects become important due to presence of a parabolic trap. 
Sec.~\ref{3D} is devoted to direct numerical simulations of the 3D GP equation and comparison of the results with predictions based on 1D models. 
The results are summarized in Conclusion. 
For the sake of convenience some relevant information on the Inverse Scattering technique (IST) and on elliptic integrals are given in Appendixes.

\section{Effective 1D equations}
\label{1Dnls}

%\subsection{Parameters of the problem}
%\label{meanfield}

As it is customary we start with the GP equation for the macroscopic wave function $\Psi({\bf r},t)$: 
\begin{equation}
\label{GPE}
i\hbar \partial_t \Psi=-\frac{\hbar^2} {2m}\Delta\Psi+V({\bf r})\Psi+g_0|\Psi|^2\Psi\, 
\end{equation}
where we use the standard notations: $g_0=4\pi\hbar^2a_s/m$, $a_s$ is the $s$-wave scattering length, and $m$ is the mass of an atom. 
The external potential is given by $V({\bf r})=V_{trap}({\bf r})+V_{latt}(x)$, where $V_{trap}({\bf r})=\frac m2 \left(\omega_{\bot}^2 r_\bot^2+\omega_0^2x^2\right)$ is a trap potential, ${\bf r}_\bot=(y,z)$, $\omega_{\bot}$ and $\omega_0$ are the transverse and axial harmonic oscillator frequencies, and $V_{latt}(x)=V_{latt}(x+\Lambda)$ is a lattice potential with the lattice constant $\Lambda$.
The wave function is normalized to the total number of atoms, ${\cal N}$: $	\int |\Psi|^2d{\bf r}={\cal N}$.

We consider a cigar-shaped trap where the transverse linear oscillator length, $a_{\bot}=\sqrt{\hbar/m\omega_{\bot}}$, is much smaller than the longitudinal one, $a_0=\sqrt{\hbar/m\omega_0}$: $a_{\bot}\ll a_0$. 
Then the condensate becomes a quasi-1D system. In literature there exist several approximations reducing the original GP equation to different effective 1D models.
The method allowing one to do that in a controlled manner, i.e. appreciating the neglected terms and using a unique well defined small parameter, is the multiple scale expansion. 
The resulting 1D models depending on relations among the parameters are summarized in the Table~\ref{tableone} (see also \cite{BK_rev}). 
\begin{table*}
\caption{Scaling, backgrounds and effective 1D equations describing matter waves in OLs. All variables in evolution equations are dimensionless (for the notations see the text).}
\begin{ruledtabular}
\begin{tabular}{clll} 
%\toprule
Case & Scaling & Background $\phi_n(x_0)$ & 1D NLS equation \\
\colrule
1& $a_{\bot}\sim \Lambda \sim\epsilon\xi\lesssim\epsilon^2a_0$ & $\approx \mbox{Bloch function}$ 
& $i\psi_t=-\frac{1}{2M}\psi_{xx}+2\sigma |\psi|^2\psi$ \\
2&$a_\bot\sim \Lambda \sim\epsilon\xi\sim \epsilon a_0$ & $=\mbox{Bloch function}$
& $i\psi_t=-\frac{1}{2M}\psi_{xx}+2\sigma |\psi|^2\psi+ \nu^2x^2\psi$ \\
3&$a_\bot\sim \epsilon \Lambda \sim\epsilon\xi\lesssim\epsilon^2a_0$ 
& $=\pi^{-1/4}e^{-\nu x_0^2/2}$
& $i\psi_t=-\psi_{xx}+2\sigma |\psi|^2\psi+U(x)\psi$\\
4& $a_\bot\sim \epsilon \Lambda \sim\epsilon\xi\sim\epsilon a_0$ 
& $\equiv 1$ & $i\psi_t=-\psi_{xx}+2\sigma |\psi|^2\psi +U(x)\psi+ \nu^2x^2\psi$ \\ 
%\botrule
\end{tabular}
\end{ruledtabular}
\label{tableone}
\end{table*}
To discuss them we introduce the mean healing length	 $\xi=(8\pi n|a_s|)^{-1/2}$, where $n\sim {\cal N}/(a_{\bot}^2a_0)$ is a mean particle density, and collect four characteristic spatial scales: $\{a_0,a_{\bot},\Lambda,\xi\}$. 
Next we introduce a small parameter of the problem, $\epsilon$, defining it through the ratio between the energy of two-body interactions, $\sim\hbar^2a_s n/m$, and the kinetic energy of the transverse excitations, $\sim\hbar^2/(ma_{\bot}^2)$: 
\begin{eqnarray}
	\label{small_pa}
\epsilon^2=\frac{a^2_\bot}{\xi^2}\sim \frac{{\cal N} a_s }{a_0}.
\end{eqnarray}
Then, one can single out four different cases~\cite{BK_rev}.

%\subsection{The models}
\label{caseone}

Let us start with the Case 1 from Table~\ref{tableone}, for which the lattice period is of order of the transverse oscillator length and much less than the healing length
%: $a_{\bot}\sim \Lambda \sim\epsilon\xi\lesssim \epsilon^2a_0$ 
(see also \cite{KonSal}).	
Introducing dimensionless independent variables $\tilde{t}=\omega_{\bot} t/2$, $\tilde{x}=x/a_{\bot}$, $\tilde{\bf r}_\bot={\bf r}_\bot/a_{\bot}$, and a renormalized wave function 
$\tilde{\Psi}=a_{\bot}|a_s|^{1/2}\Psi$, we rewrite Eq.~(\ref{GPE}) in the dimensionless form
\begin{eqnarray}
\label{GPE1}
\begin{array}{l}
\displaystyle{
i\partial_{\tilde{t}}\tilde{\Psi} =\left({\cal L}_\bot+{\cal L}_0 +8\pi\sigma|\tilde{\Psi}|^2\right)\tilde{\Psi},
}
\\
\displaystyle{
{\cal L}_\bot=-\Delta_\bot + \tilde{r}^2_\bot,\quad {\cal L}_0=-\partial_{\tilde{x}}^2 +\nu^2 \tilde{x}^2+\kappa^2U(\kappa \tilde{x}).
}
%\nonumber
\end{array}
\end{eqnarray}
Hereafter $\sigma={\rm sign}(a_s)$, $\nu=\omega_0/\omega_\bot$ is the aspect ratio of the trap, $U(\kappa\tilde{x})\equiv V_{latt}(x)/E_R$ is a dimensionless periodic potential which is varying on a unit scale, $E_R=\hbar^2\pi^2/(2\Lambda^2 m)$ is the recoil energy, and 
%\begin{eqnarray}
%\label{gamma1}
$
\kappa^2=\frac{E_R}{\hbar\omega_\bot}\sim\frac{a_\bot^2}{\Lambda^2}.
$
%\end{eqnarray}

Next we consider linear eigenvalue problems 
%associated to (\ref{GPE1}), (\ref{operators})
\begin{eqnarray}
	\label{lin_prob}
 {\cal L}_0\, \phi_{nq}(\tilde{x})&=&{\cal E}_{nq}\, \phi_{nq}(\tx),
 \\
{\cal L}_\bot\, \xi_{l_1l_2}(\tilde{{\bf r}}_\bot)&=&{\cal E}_{l_1l_2}\, \xi_{l_1l_2} (\tilde{{\bf r}}_\bot). 
%\nonumber
\end{eqnarray}
Hereafter $n$ and $q$ stand for a band number and a wave vector inside the first Brillouin zone (BZ), when it is relevant (cases 1 and 2 in Table~\ref{tableone}) and the index $q$ must be dropped otherwise (cases 3 and 4 in Table~\ref{tableone}). Indexes $l_{1,2}$ stand for two transverse quantum numbers. 
When $\nu\lesssim \epsilon$, $\phi_{nq}(\tilde{x})$ can be approximated by a Bloch function of the respective periodic potential. 
We restrict the consideration to the Bloch states bordering the edge of the BZ where $q=q_0= \pi/\Lambda$, and respectively use the abbreviated notations $\phi_n(\tilde{x})\equiv \phi_{nq_0}(\tilde{x})$ and ${\cal E}_n\equiv {\cal E}_{nq_0}$. 
We also consider only the evolution of the background state ($l_1=l_2=0$) of the transverse component: $\xi_{00}(\tilde{{\bf r}}_\bot)=\exp(-\tr^2/2)/\sqrt{\pi}$, and ${\cal E}_{00}=2$. 

Now we look for a solution of Eq.(\ref{GPE1}) in a form 
$
\tilde{\Psi}=\epsilon\psi_1+\epsilon^2\psi_2+\cdots,
$
where $\psi_j$ ($j=1,2,\ldots$) are functions of $\tilde{{\bf r}}_\bot$ and of scaled independent variables $t_\alpha=\epsilon^\alpha \tilde{t}$ and $x_\alpha=\epsilon^\alpha \tilde{x}$. 
Then the leading order of the solution can be searched in the form of the modulated ground state 
\begin{eqnarray}
\label{gr_st}
\psi_1=\frac{1}{\sqrt{\pi}} {\cal A}(x_1,t_2) e^{-i({\cal E}_{n}+2)t_0}e^{-\tr^2/2}\phi_{n}(x_0),	
\end{eqnarray}
where ${\cal A}(x_1,t_2)$ is a slowly varying envelope amplitude in which by convention we indicate only the most rapid variables.

The next steps are standard for the multiple-scale expansion: 
substituting $\tilde{\Psi}$ in (\ref{GPE1}) one resolves the equations appearing in the three lowest orders of $\epsilon$ (see e.g. \cite{BK_rev,KonSal} for details) and subsequently eliminates secular terms by imposing requirements on the amplitude ${\cal A}(x_1,t_2)$.
For the case at hand this results in equation
\begin{eqnarray}
\label{nls1D}
i \partial_t \psi =-\frac{1}{2M}\partial^2_x \psi+ 2\sigma|\psi|^2\psi,	
\end{eqnarray}
where $\psi=\sqrt{\chi/2}{\cal A}$ with $ 
\chi=2\int_0^{\Lambda} |\phi_{n}(x)|^4dx$, and for the sake of brevity we made substitutions $t_2\to t$ and $x_1\to x$. 
The effective mass is obtained in the formal limit $a_0/\Lambda\to \infty$: $M^{-1}=\left(d^2{\cal E}_{nq}/dq^2\right)_{q=q_0}$ \cite{KonSal}. 
 
When the condensate size in the axial direction is not large enough and ``boundary'' effects cannot be neglected (Case 2 from Table \ref{tableone}),
%: where $a_\bot\sim \Lambda \sim \epsilon\xi\sim\epsilon a_0$, 
the linear spectral problem must be modified. 
Since now $a_\bot/a_0\sim\epsilon$ and thus $\nu^2x^2=\epsilon^2\tilde{\nu}^2x_1^2$ with $\nu=\epsilon^2\tilde{\nu}$ and $\tilde{\nu}=O(1)$, the problem (\ref{GPE1}) is to be redefined as follows
\begin{eqnarray}
\label{GPE2}
\begin{array}{l}
\displaystyle{
i\partial_{\tilde{t}} \tilde{\Psi} =\left({\cal L}_\bot+{\cal L}_0 +\epsilon^2\tilde{\nu}^2 (\epsilon \tx)^2
+8\pi\sigma|\tilde{\Psi}|^2\right)\tilde{\Psi},
}
\\
\displaystyle{ 
{\cal L}_{\bot}=- \Delta_\bot + \tr^2, \quad 
{\cal L}_0=-\partial^2_{\tx} + \kappa^2U(\kappa\tx).
}
\end{array}
\end{eqnarray}
Then the resulting 1D equation reads 
\begin{eqnarray}
\label{nls1dcase2}
i\partial_t\psi=-\frac{1}{2M}\partial^2_{x}\psi + \nu^2x^2\psi +2\sigma |\psi|^2\psi 	
\end{eqnarray}
where for the sake of simplicity $\tilde{\nu}$ is substituted by $\nu$ and as before ($\psi$, $t$, $x$) stand for ($\sqrt{\chi/2}{\cal A}$, $t_2$, $x_1$).
 
If the period of the potential is large compared with the healing length 
%i.e. $a_\bot\sim \epsilon \Lambda \sim\epsilon\xi\lesssim\epsilon^2a_0 $ 
(Case 3 from Table~\ref{tableone}), one has $\kappa=\epsilon\tilde{\kappa}$, where $\tilde{\kappa}$ is of order one. 
Then $\kappa^2U(\kappa \tx)\equiv \epsilon^2\tilde{U}(x_1)$ where $\tilde{U}(x_1)\equiv \tilde{\kappa}^2U(\tilde{\kappa} x_1)$ and system (\ref{GPE1}) acquires the form
\begin{eqnarray}
\label{GPE3}
\begin{array}{l}
\displaystyle{
i\partial_{\tilde{t}} \tilde{\Psi} =\left({\cal L}_\bot+{\cal L}_0 +\epsilon^2 \tilde{U}(\epsilon \tx)
+8\pi\sigma|\tilde{\Psi}|^2\right)\tilde{\Psi},
}
\\
\displaystyle{
{\cal L}_\bot=- \Delta_\bot + \tr^2, \qquad 
{\cal L}_0=-\partial_{\tx}^2+\nu^2 \tx^2.
} 
\end{array}
\end{eqnarray}
The eigenvalue problem for ${\cal L}_0$ is now solved explicitly: $\phi_0(x_0)=\pi^{-1/4}\exp(-\nu x_0^2/2)$, and the final (i.e. already in dimensionless units) 1D equation reads
 \begin{eqnarray}
\label{case3nls} 
i\partial_t\psi=-\partial^2_x\psi+ U(x)\psi +2\sigma |\psi|^2\psi.
\end{eqnarray}
Here $\tilde{U}(x_1)$ is substituted by $U(x)$.

Finally, if one considers not too long condensate and a lattice having a period of order of the mean healing length (Case 4 in Table~\ref{tableone}),
the corresponding problem is reformulated now as follows 
\begin{eqnarray}
\label{GPE4}
\begin{array}{l}
\displaystyle{
i\partial_{\tilde{t}}\tilde{\Psi} =\left({\cal L}_\bot+{\cal L}_0 
+ \epsilon^4\tilde{\nu}^2 \tx^2 + \epsilon^2\tilde{U}(\epsilon \tx) +8\pi\sigma |\tilde{\Psi}|^2\right)\tilde{\Psi},
}
\\
\displaystyle{
{\cal L}_\bot =- \Delta_\bot + \tr^2, \qquad 
{\cal L}_0=-\partial_{\tx}^2.
}
\end{array}
\end{eqnarray}
The ground state as a solution of the linear eigenvalue problems (\ref{lin_prob}) becomes $\phi_n\equiv 1$ and the 1D model reads 
\begin{equation}
\label{nls1_case4} 
i\partial_t\psi=-\partial^2_x\psi+ \nu^2x^2\psi+U(x)\psi+2\sigma|\psi|^2\psi
\end{equation}
where as before the tildes are suppressed.

\section{NLS equation with time dependent nonlinearity. Adiabatic theory}
\label{NLS}

Cases 1 and 3 from Table~\ref{tableone} allow rather complete description within the framework of the perturbation theory, while the Cases 2 and 4 where the parabolic potential is explicitly included in the evolution equation for the slowly varying amplitude, have to be studied by means of numerical simulations. In the present section we reproduce the main results of \cite{prl90AKKB} providing detail analysis of the asymptotics of the solutions.

\subsection{Stationary periodic waves}

Let us recall known formulas for periodic solutions, $\psi(x,t)=\psi(x+L,t)$, of the NLS equation (\ref{nls1D}), where without restriction of generality we put $|M|=1/2$ and define $\tilde{\sigma}=\sigma\,$sign$(M)$: 
\begin{equation}
\label{nls1} 
i\psi_t=-\psi_{xx}+2\tsigma|\psi|^2\psi.
\end{equation}
(In the physical units the period is recovered to be $a_\bot L/\epsilon$, and thus, in accordance with the scaling of the Case 1, $a_\bot L/\epsilon \sim \xi\gg \Lambda$.) 
The simplest stationary solutions of such type can be obtained in a form of a standing wave 
\begin{eqnarray}
	\label{subs1}
	\psi(x,t)=e^{i\omega t}u(x)
\end{eqnarray}
where $u(x)$ is a real periodic function, $u(x)=u(x+L)$, and $\omega$ is a real frequency. 
In this case (\ref{nls1}) is reduced to the ordinary differential equation 
$-u_{xx}+\omega u +2\tsigma u^3=0$ which yields
\begin{eqnarray}
	\label{solut1}
	x-x_0=\int_{0}^{u}\frac{du}{\sqrt{C+\omega u^2+\tsigma u^4}}.
\end{eqnarray}
Here $x_0$ and $C$ are the integration constants. 
As far as the stationary solution $u(x)$ is found, a periodic wave moving with a constant velocity $V$ is obtained as
\begin{eqnarray}
	\label{subs2}
	\psi(x,t)=\exp\left[i\left(\omega-\frac{V^2}{4}\right) t+i\frac{V}{2}x\right]u(x-Vt).
\end{eqnarray}
 
Next steps depend on a particular choice of the parameters $C$ and $\omega$. It turns out, however, that for our purposes another parameterization, provided by the inverse scattering technique (IST), is more convenient. 

Such parameterization is introduced in Appendix A and is summarized in Table~\ref{simplest} where the well known analytical expressions for periodic solutions are given (we use the standard notations for the Jacobi elliptic functions pq$(x,m)$ with p=c, s, d and q=n, \cite{Abramowitz}).
\begin{table*}[ht]
\caption{The simplest static periodic solutions.}
\begin{ruledtabular}
\begin{tabular}{ccccc} 
%\toprule
$\tsigma$ & periodic solution & $m$ & limit $m\to 0$ & limit $m\to 1$ \cr
\hline
\noalign{\smallskip}\hline\noalign{\smallskip}
$1$ & $\displaystyle{\psi_{\sn}=e^{i(\xi_1^2+\xi_2^2)t/2}\frac{\xi_1-\xi_2}{2}\,\sn\left(\frac{\xi_1+\xi_2}{2}x, m \right)}$ &
 $\displaystyle{\frac{(\xi_1-\xi_2)^2}{(\xi_1+\xi_2)^2}}$ & 
 $\displaystyle{e^{i\xi^2 t}\frac{\Delta\xi}{2}\,\sin(\xi x)}$ 
& $\displaystyle{\frac{\xi}{2}e^{-i\xi^2t/2}\tanh\left(\frac{\xi}{2}x\right)} $
\cr
\noalign{\smallskip}\hline\noalign{\smallskip}
$-1$ & $\displaystyle{\psi_{\dn}=e^{-i(\eta_1^2+\eta_2^2)t/2}
\frac{\eta_1+\eta_2}{2}\,\dn\left(\frac{\eta_1+\eta_2}{2}x, m \right)}$ & $\displaystyle{\frac{4\eta_1\eta_2}{(\eta_1+\eta_2)^2}}$ &
$\displaystyle{ \eta e^{-i\eta^2t} }$
&$\displaystyle{\eta e^{i\eta^2t}\mbox{sech}(\eta x)}$\cr
\noalign{\smallskip}\hline\noalign{\smallskip}
$-1$ & $\displaystyle{\psi_{\cn}=e^{i(\xi^2-\eta^2)t/2}\eta\,\cn\left(\sqrt{\eta^2+\xi^2}x, m\right)}$
& $\displaystyle{\frac{\eta^2}{\xi^2+\eta^2}}$ &
$\displaystyle{e^{-i\eta^2t/2}\eta\,\cos (\eta x)}$
&$\displaystyle{\eta e^{i\eta^2t}\mbox{sech}(\eta x)}$ 
\\ 
%\botrule
\end{tabular}
\label{simplest}
\end{ruledtabular}
\end{table*}

\subsection{BEC with a varying scattering length}
\label{Feshbach}

As it is explained in Appendix A, controlled change of a shape of a periodic wave can be described by motion of the roots $\lambda_j$ on the complex plane. 
In particular, to transform a periodic wave into a soliton one has to find a way to move accordingly the branching points $\lambda_j$. 
It turns out that the limiting transitions shown in Table~\ref{tableone} and in Fig.\ref{figone} cannot be reached in practice because they require change of the boundary conditions of the problem and thus change of the number of atoms involved. 
That is why instead of looking for a possibility of transforming a periodic wave in a single soliton one can to explore a possibility of its transformation in a sequence (we call it train) of solitons equally spaced and thus preserving the period of the wave.
This can be done by perturbing the evolution equation. This has to be done  adiabatically in order to do not destroy the regular structure of the periodic wave. 
In particular, one can consider change of the effective mass by means of variation of either the laser field intensity or geometry of laser beams, or change of the nonlinearity. The latter can be achieved by means of the FR affecting the scattering length and controlled by an external magnetic filed $B(t)$ varying with time. In that case a model simulating the process can be written as follows \cite{14}:
\begin{equation}
\label{SL}
a_{s}(t) = a_s(0)g(t),\quad g(t)=\frac{\Delta B (0)(\Delta B(t)+\Delta_{res})}{\Delta B (t)(\Delta B(0)+\Delta_{res})}.
\end{equation}
Here $a_s(0)$ is the initial value of the scattering length, $B(0)=B_0+\Delta B(0)$ is the initial field, $\Delta B(t)$ is a deviation from the resonant value $B_{0}$, and $\Delta_{res}$ is the width of the resonance.
 
Assuming that dependence of $g(t)$ on time is slow enough (i.e. is governed by $\epsilon^2 \tilde{t}$) one can repeat the arguments of Sec.~\ref{1Dnls}, substituting $g_0$ by $\tilde{g}_0=g_0\cdot g(t)$, what results in modification of each of the evolution equations (\ref{nls1D}) [respectively (\ref{nls1})], (\ref{nls1dcase2}), (\ref{case3nls}), and (\ref{nls1_case4}). In particular (\ref{nls1}) now reads 
\begin{equation}\label{NLS-F}
i\psi_t+ \psi_{xx}- 2\tsigma g(t)|\psi|^2\psi=0.
\end{equation} 
 
In what follows we consider only the case when $g(t)$ is a positive function, i.e. where the FR does not change a type of the two-body interaction.
Then, in order to study evolution of the condensate within the framework of the model (\ref{NLS-F}) we notice that substitution
\begin{equation}
\label{u-v}
 \psi(x,t)=\frac{v(x,t)}{\sqrt{g(t)}}=
 e^{-\zeta(t)/2} v(x,\zeta) 
\end{equation}
where
\begin{eqnarray}
\label{z}
 \zeta=\zeta (t)=2\int_0^t\gamma(t') dt'=\ln g(t), \qquad \zeta(0)=0
\end{eqnarray}
transforms (\ref{NLS-F}) into a dissipative NLS equation
\begin{equation}
\label{modNLS}
 iv_t+v_{xx}-2\tsigma |v|^2v=i\gamma v
\end{equation}
where
\begin{equation}\label{gamma}
\gamma(t)=\frac{1}{2g(t)}\frac{dg(t)}{dt}=\frac 12 \frac{d\zeta}{dt} .
\end{equation}
For slowly varying $g(t)$ the right hand side of Eq.~(\ref{modNLS})
can be considered as a small perturbation: $|\gamma(t)|\ll 1$. It is to be emphasized here that slow dependence means slow in terms of slow time $\epsilon^2\tilde{t}$, what is satisfied, for example by the function $g$ depending on $\epsilon^{5/2}\tilde{t}$.
If the nonlinearity increases ($\gamma(t)>0$, the case we will deal with),  the perturbation describes growth, otherwise, when the nonlinearity is decaying ($\gamma(t)<0$) the perturbation describes dissipation.

Since $g(0)=1$ it
follows from Eq.~(\ref{u-v}), that $v(x,0)=\psi(x,0)$. 

\subsection{Adiabatic approximation}

Under influence of the dissipative perturbation a wave shape is changed what can be described by variations of the parameters $\lambda_j$ assuming them to be slow functions of time $t$ (the so-called {\em adiabatic approximation}). Equations which govern their evolution can be derived by the following simple method \cite{prl90AKKB,VVKBook}.

First of all we recall that in all the examples considered in this section the waves are two-parametric and hence we have to obtain two equations of the adiabatic approximation. 
Since the coefficients in the dissipative model (\ref{modNLS}) are supposed to be independent on $x$, the perturbation does not result in change of the period $L=L(\la_1,\la_2)$ of the nonlinear wave what can be expressed as
\begin{equation}
\label{wl}
 \frac{d}{d\zeta}L(\la_1,\la_2)=0.
\end{equation}
Here we temporarily use $\la_{1,2}$ for two parameters of a periodic solution: thus $\la_{1,2}=\xi_{1,2}$ for the sn-wave, $\lambda_{1,2}=\eta_{1,2}$ for the dn-wave, and $\lambda_{1}=\eta$ and $\la_{2}=\xi$ for the cn-wave (see Appendix A).

Next, it is a straightforward algebra to show that the integral 
\begin{eqnarray}
N=\int_{0}^{L} |v|^2dx
\end{eqnarray}
of a periodic solution of (\ref{modNLS}) (below it will be referred to as a number of particles, assuming that there is no confusion with a real number of particles given by ${\cal N}$, see Sec.~\ref{1Dnls}) evolves according to
\begin{equation}\label{Neq}
 \frac{dN(\la_1,\la_2)}{d\zeta}= N(\la_1,\la_2).
\end{equation}
 
Then, if the expressions for $L$ and $N$ in terms of $\la_{1,2}$ are
known, Eqs.~(\ref{wl}) and (\ref{Neq}) reduce to a system of two first order differential equations for $\la_{1,2}$.

\subsection{sn-wave in a BEC with $a_sM>0$}
\label{subsec:sn-period}

We start with the adiabatic dynamics of a periodic matter wave in a BEC with a positive scattering length and look for a solution of (\ref{modNLS}) in a form of the sn-wave given in the first row of Table~\ref{simplest}. 
Then
\begin{eqnarray}
	\label{period_sn}
	L=\frac{8{\rm K}(m)}{\xi_1+\xi_2},\qquad N=2(\xi_1+\xi_2)[{\rm K}(m)-{\rm E}(m)]
\end{eqnarray}
where ${\rm K}(m)$ and ${\rm E}(m)$ are standard notation for the complete elliptic integrals of the first and second kinds, respectively~\cite{Abramowitz,Lawden} and $m$ is defined through $\xi_j$ in Table~\ref{simplest}. 
Now, considering $\xi_j$ as functions of time, or more precisely of $\zeta=\zeta(t)$, and substituting (\ref{period_sn}) into (\ref{wl}) and (\ref{Neq}) we obtain
\begin{eqnarray}
\label{sn-syst}
 \begin{array}{l}\displaystyle{
 \frac{d\xi_1}{d\zeta}=\frac{\xi_1(\xi_1+\xi_2)[{\rm K}(m)-{\rm E}(m)]}
 {2\xi_1{\rm K}(m)-(\xi_1+\xi_2){\rm E}(m)}},\\
 \displaystyle{
 \frac{d\xi_2}{d\zeta}=\frac{\xi_2(\xi_1+\xi_2)[{\rm K}(m)-{\rm E}(m)]}
 {2\xi_2{\rm K}(m)-(\xi_1+\xi_2){\rm E}(m)}}.
 \end{array}
\end{eqnarray}

For the sake of definiteness we assume that the parameters $\xi_j$ are chosen such that initially $\xi_{01}>\xi_{02}$ as in Fig.\ref{figone}a (hereafter $\xi_{0j}=\xi_j(t=0)$). 
Then one immediately concludes that $d\xi_1/d\zeta>0$, i.e. $\xi_1$ is a monotonically growing function of time. 
One can also show (see Appendix~\ref{elliptic}) that $\xi_2$ is a monotonically decreasing function and thus $m\to 1$ as time goes to infinity. 
To describe the respective limiting behavior we employ  asymptotics of the elliptic integrals (\ref{int-asymp}), and reduce (\ref{sn-syst}) in the leading order (when $\xi_1\to \infty$ and $\xi_2\to 0$) to 
\begin{eqnarray}
	\label{asymp-sn-eq}
	\frac{d\xi_1}{d\zeta}=\frac{\xi_1}{2},\qquad \frac{d\xi_2}{d\zeta}=-\frac{\xi_2 }{2}
	\ln\frac{\xi_1}{\xi_2}.
\end{eqnarray}
The first of these equations is trivially solved. To solve the second one we observe that from (\ref{asymp-sn-eq}) it follows that
\begin{eqnarray}
\label{univ-sn}
	\frac{d\xi_2}{d\xi_1}=-\frac{\xi_2}{\xi_1}\ln\frac{\xi_1}{\xi_2}.
\end{eqnarray}
This equation is also readily solved giving
\begin{eqnarray*}
	\xi_2=\exp(1+\ln\xi_1+C_0\xi_1)\sim e^{C_0\xi_1}.
\end{eqnarray*}
Here $C_0$ is an integration constant, which can be found from the (asymptotic) condition for conservation of the period computed from (\ref{period_sn}): 
%\begin{eqnarray}
$
	L\sim\frac{4}{\xi_1}\ln\frac{\xi_1}{\xi_2}\sim -4C_0.
$
%\end{eqnarray}
Finally we arrive at the asymptotics:
\begin{eqnarray}
	\label{sn-asymp}
	\xi_1\sim e^{\zeta/2},\qquad \xi_2\sim\xi_1e^{-L\xi_1/4} .
\end{eqnarray}

Thus, one observes a scenario different from one shown in Fig.\ref{figone}a,b (see the example in Fig.\ref{sn-pot}b, below). 
Now $\xi_1$ goes to the infinity and we obtain a ``train of dark solitons''. This is related to the fact that the perturbation does not affect the period of the nonlinear wave, while in order to obtain a single dark soliton the period should go to the infinity in the process of the wave deformation. Meantime, the shape of the train of dark solitons indeed approximates to a shape of a set of static dark solitons, what occurs due to extremely rapid convergence of $\xi_2$ to zero, what is illustrated in inset in Fig.\ref{sn-pot}b.

\subsection{dn-wave in a BEC with $a_sM<0$}

Let us now turn to the case $\tsigma=-1$, and consider deformations of the periodic dn-wave, given by the second row in Table~\ref{simplest}, resulting in a train of bright solitons (the counterpart of the transformation shown in Figs.\ref{figone}c,d). 
The period $L$ and the number of particles $N$ are now given by
\begin{eqnarray}
	\label{period-dn}
	L=\frac{4{\rm K}(m)}{\eta_1+\eta_2},\qquad N=(\eta_1+\eta_2){\rm E}(m).
\end{eqnarray}
Assuming that parameters $\eta_j$ depend on time and substituting (\ref{period-dn}) in (\ref{wl}) and (\ref{Neq}) we obtain a system of the equations of the adiabatic approximation for $\eta_1$ and $\eta_2$:
\begin{equation}\label{dn-syst}
 \begin{array}{l}\displaystyle{
 \frac{d\eta_1}{d\zeta}=\frac{\eta_1(\eta_1+\eta_2){\rm E}(m)}
 {(\eta_1+\eta_2){\rm E}(m)+(\eta_1-\eta_2){\rm K}(m)}},\\
 \displaystyle{
 \frac{d\eta_2}{d\zeta}=\frac{\eta_2(\eta_1+\eta_2){\rm E}(m)}
 {(\eta_1+\eta_2){\rm E}(m)-(\eta_1-\eta_2){\rm K}(m)}}
 \end{array}
\end{equation}
where $m$ is given in Table~\ref{simplest}, $\zeta$ is defined by Eq.~(\ref{z}), and for the sake of definiteness it is supposed that $\eta_2>\eta_1$.

Like in the previous subsection one can show that the both $\eta_j$ are monotonically increasing functions (see Appendix~\ref{elliptic}). Moreover, it is evident that $\eta_2$ growth more rapidly than $\eta_1$. To consider the limit $|\eta_1-\eta_2|\to 0$, i.e.  $m\to 1$, we introduce $\eta=(\eta_1+\eta_2)/2$ and $\Delta\eta=(\eta_2-\eta_1)/2$. 
Then using (\ref{int-asymp}) we obtain from (\ref{dn-syst}):
\begin{eqnarray}
	\label{dn-asymp-eq}
	\frac{d\eta}{d\zeta}=\eta,\qquad \frac{d\Delta\eta}{d\zeta}=-\Delta\eta 
	\ln\frac{\eta}{\Delta\eta}.
\end{eqnarray}
Comparing these equations with (\ref{asymp-sn-eq}) one observes the {\em universality} of the limiting behavior. 
Thus using (\ref{sn-asymp}) one immediately writes down the asymptotic formulas (as in the previous case the constant is found from the conservation of the wave period) 
\begin{eqnarray}
	\label{dn-asymp}
	\eta\sim e^{\zeta},\qquad \Delta\eta\sim\eta e^{-L\eta/4}. 
\end{eqnarray}

A peculiarity of the behavior of the dn-wave parameters in the limit $m\to 1$ is that both $\eta_1$ and $\eta_2$ tend to infinity (see the inset in Fig.\ref{dn-pot}a, below), what is consistent with the fact that keeping the wave period to be a constant one cannot obtain a single bright soliton. 
Meantime, as we could expect the difference $|\eta_1-\eta_2|$ rapidly goes to zero, what corresponds to appearance of a ``train of bright solitons'' (see the example in Fig.\ref{dn-pot}a). 
 
\subsection{cn-wave in a BEC with $a_sM<0$}

Finally, we consider the case of the cn-wave (the third line of Table~\ref{simplest}) in a BEC with a negative scattering length. 
The period and the number of particles are now given by
\begin{eqnarray}
	\label{period-cn}
	L=\frac{4{\rm K}(m)}{\sqrt{\xi^2+\eta^2}},\qquad N=4\sqrt{\xi^2+\eta^2}
	{\rm E}(m)-\xi^2L.
\end{eqnarray}
Substitution of these expressions into Eqs.~(\ref{wl}) and (\ref{Neq}) yields 
\begin{equation}\label{cn-syst}
 \begin{array}{l}\displaystyle{
 \frac{d\eta}{d\zeta}=\frac{[(\eta^2+\xi^2){\rm E}(m)-\xi^2{\rm K}(m)]{\rm E}(m)\eta}
 {\eta^2{\rm E}^2(m)+\xi^2[{\rm K}(m)-{\rm E}(m)]^2},}\\
 \displaystyle{
 \frac{d\xi}{d\zeta}=\frac{[\xi^2{\rm K}(m)-(\eta^2+\xi^2){\rm E}(m)][{\rm K}(m)-{\rm E}(m)]\xi}
 {\eta^2{\rm E}^2(m)+\xi^2[{\rm K}(m)-{\rm E}(m)]^2}}.
 \end{array}
\end{equation}

Now $\eta$ and $\xi$ are respectively increasing and decreasing functions.
As in the two previous examples analytical treatment can be given for the limiting case, which is now $\xi\to 0$ and $\eta\to \infty$. Using (\ref{int-asymp}) it is a straightforward algebra to ensure that in the case at hand we arrive at already familiar equations    
\begin{eqnarray}
	\label{cn-asym-eq}
	\frac{d\eta}{d\zeta}=\eta\,,\qquad \frac{d\xi}{d\zeta}=-\xi\ln\frac{\eta}{\xi}
\end{eqnarray}
describing the {\em universal} law of the limiting transition [c.f. (\ref{sn-asymp}) and (\ref{dn-asymp})]
\begin{eqnarray}
	\label{cn-asymp}
	\eta\sim e^{\zeta}, \qquad
	\xi\sim\eta e^{-L\eta/4}\,. 
\end{eqnarray}
Dynamics of respective parameters $\eta$ and $\xi$ on time is shown in the inset in Fig.\ref{cn-pot}b.
 
\section{BEC subject to FR in a periodic potential having large period } 
\label{ini}

\subsection{Stationary solution and adiabatic approximation}
 
The Case 1 from Table~\ref{tableone} is the simplest one since it deals with the integrable NLS equation (\ref{nls1D}). 
It turns out that the Case 3 from Table~\ref{tableone} with varying nonlinearity can also be treated within the framework of the adiabatic perturbation theory, due to the fact that some solutions of the NLS equation with a constant nonlinearity are known explicitly. 
They were considered in details in Ref.~\cite{24} and are partially summarized in the present subsection.

To this end we notice that ansatz (\ref{subs1}) now results in the stationary equation 
\begin{eqnarray}
\label{differeq}
u_{xx}=(\omega +U(x) + 2\tsigma u^2) u,
\end{eqnarray} 
which allows one to construct a lattice potential $U(x)$, in an explicit form, when a stationary periodic solution of the NLS equation (\ref{nls1}), is given a priori. 
Indeed, one can verify, that (\ref{subs1}) with $u(x)=\sqrt{\chi}u_0(x)$, $\chi>0$, solves (\ref{case3nls}) with a periodic potential $U(x)=2\tsigma (1-\chi)|u_0(x)|^2$ if $\psi=e^{i\omega t} u_0(x)$ solves (\ref{nls1}). 
Alternatively, from Eq. (\ref{differeq}) one can find a lattice potential $U(x)$ for a given wave field $u(x)$, subject to the requirement of regular behavior of $u_{xx}/u$: $U(x)=\frac{u_{xx}}{u}-\omega-2\tsigma u^2$. 
By any of the above methods, one can verify that the Jacobi elliptic functions pq$(\kappa x,m)$, where p=c,s or d, and ${\rm q=n}$ [more generally one can consider $u_{\rm pq}(x)=\left(A\, {\rm pq}^\beta(\kappa x,m)+B\right)^\alpha$ with $B> |A|$, $\alpha$ and $\beta$ being real integers], subject to proper choice of $\tsigma$, give the lattice potential 
\begin{equation}
\label{Vlatt}
U(x)=U_0\sn^2(\kappa x,m),
\end{equation}
where $U_0$ is a constant and we returned to the parameterization given by $\kappa$ and $m$. (Although one can continue using $\lambda_j$, the latter now loose the meaning they had in the context of the IST.)
The respective solutions are summarized in the Table~\ref{tabletwo}.
%Table 3%%%%%%%%%%%%%%%%%%%%%%%%%%%%%%%%%%%%%%%%%%%%
%%%%%%%%%%%%%%%%%%%%%%%%%%%%%%%%%%%%%%%%%%%%%
\begin{table}[ht]
\caption{Solutions of the type $u_{{\rm pq}}(x,t)=\sqrt{A}\, {\rm pq}(\kappa x,m)$. The parameters in the second column have to be chosen to provide $A>0$.}
\begin{tabular}{ccc} \toprule
pq & $A$ & $\omega$ \cr
\hline
\noalign{\smallskip}\hline\noalign{\smallskip}
$\sn$\hspace{0.5cm} & $(\kappa^2m-U_0/2)\tsigma$\hspace{0.5cm} & $-\kappa^2(1+m)$\cr
\noalign{\smallskip}\hline\noalign{\smallskip}
$\cn$\hspace{0.5cm} & $(U_0/2-\kappa^2m)\tsigma$\hspace{0.5cm} & $\kappa^2(2m-1)-U_0$ \cr
\noalign{\smallskip}\hline\noalign{\smallskip}
$\dn$\hspace{0.5cm} & $(U_0/2m-\kappa^2)\tsigma$\hspace{0.5cm} & $\kappa^2(2-m)-U_0/m$ \\ \botrule
\end{tabular}
\label{tabletwo}
\end{table}

If the nonlinearity depends on time, we again make use of the substitution (\ref{u-v}) leading to the equation [c.f. (\ref{modNLS})]
\begin{equation}
\label{modNLS1}
 iv_t+v_{xx}-U_0\sn^2(\kappa x,m)v-2\tsigma |v|^2v=i\gamma v.
\end{equation}
As above the right hand side of Eq.~(\ref{modNLS1})
is considered as a small perturbation: $|\gamma(t)|\ll 1$. Then, in accordance with the adiabatic approximation, the parameters $\kappa$ and $m$ will be considered as slow functions of time $t$, and respectively will be designated as $\tkappa$ and $\tm$. The equations of their evolution are obtained in the same way as we obtained above the equations for the dependence of $\lambda_{1,2}$ on time. 
More specifically, they are Eqs.~(\ref{wl}), (\ref{Neq}), where the period now coincides with the period of the periodic potential and  $L$ and $N$ are expressed through $\tkappa$ and $\tm$: $L=L(\tkappa,\tm)$, $N=N(\tkappa,\tm)$. 
The time-dependent parameters must satisfy initial conditions as follows
\begin{equation}
	\label{init_mk}
	\tm(0)=m,\qquad \tkappa (0)=\kappa
\end{equation}
where respective links between $m$ and $\kappa$ are given in the Table~\ref{tabletwo}.
 
\subsection{BEC with a positive scattering length: sn-wave}
\label{sec:sn-latt}
 
Let us start with adiabatic dynamics of a periodic matter wave in a BEC with a positive scattering length ($\tsigma=1$) and with $2\tkappa^2 \tm>U_0$, loaded in OL (\ref{Vlatt}). 
Substituting solution for the sn-wave (the first row of Table~\ref{tabletwo}):
\begin{eqnarray}
	\label{sn_wave_pot}
v_{\sn}=\sqrt{\tkappa^2\tm-\frac{U_0}{2}}e^{-i\tkappa^2(1+\tm)t}\sn (\tkappa x, \tm)
\end{eqnarray}
into (\ref{modNLS1}) one obtains: 
\begin{eqnarray}
\label{sn-lat-L}
L=\frac{4{\rm K}(\tm)}{\tkappa},
\quad
N=\frac{4\tkappa^2\tm-2U_0}{\tkappa \tm}\left[{\rm K}(\tm)-{\rm E}(\tm)\right].
\end{eqnarray}

\begin{figure}[ht]
%\vspace{0.5 true cm}
\epsfig{file=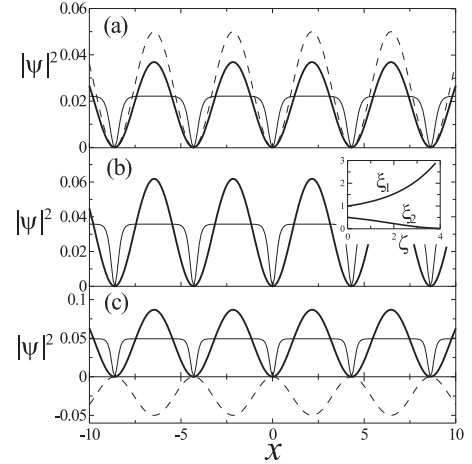,width=6cm}
%\vspace{0.5 true cm}
\caption{Examples of sn-wave solutions at $\zeta=0$ (solid thick lines) and $\zeta=\zeta_{fin}$ (solid thin lines) in the periodic potential (\ref{Vlatt}) (dashed lines) for (a) $U_0=0.05$, and $\zeta_{fin}=5$; (b) $U_0=0$, $\zeta_{fin}=6$, and (c) $U_0=-0.05$, $\zeta_{fin}=6$. 
The inset shows dynamics of the parameters $\xi_1$ and $\xi_2$, which connection with $\tilde{m}$ and $\tilde{\kappa}$ is given in Table~\ref{simplest}. 
In all the cases $m=0.11$ and $\kappa=0.75$.}
\label{sn-pot} 
\end{figure}

Now one can distinguish two different situations, where maxima of the atomic density are located (i) in maxima ($U_0>0$) and (ii) in minima ($U_0<0$) of the periodic potential (examples are shown in Fig.\ref{sn-pot} (a) and (c), respectively). 
It is interesting to notice that, although the last situation admits the limit $\tm\to 0$, where the distribution is transformed into a sine-wave: $v_{\sn}\to\sqrt{|U_0|/2}e^{-i\tkappa^2 t}\sin(\tkappa x)$, it does not allow transition to the linear limit in the sense of a small amplitude, the latter being possible only if $U_0 > 0$, corresponding to $\tm\to U_0/(2\tkappa^2)$. 

For the next step we substitute (\ref{sn-lat-L}) into (\ref{wl}) and (\ref{Neq}), what gives 
\begin{eqnarray} \label{sn-pot-ad1} 
 \begin{array}{l}\displaystyle{
\frac{d\tkappa}{d\zeta}=\frac{\tkappa(2\tkappa^2\tm-U_0)}{\Delta_{\sn}} }\cr
\displaystyle{\qquad\qquad\times\left[{\rm E}(\tm)-{\rm K}(\tm)\right]\left[{\rm E}(\tm)-\tmp{\rm K}(\tm)\right],}\cr
	\displaystyle{
\frac{d\tm}{d\zeta}=\frac{2\tm \tmp(2\tkappa^2\tm-U_0)}{\Delta_{\sn}}{\rm K}(\tm)\left[{\rm E}(\tm)-{\rm K}(\tm)\right].}
 \end{array}
\end{eqnarray}
Here standard notation $\tmp(\zeta)\equiv 1-\tm(\zeta)$ is used and
\begin{eqnarray*}
	\Delta_{\sn}&=&U_0\left[{\rm E}^2(\tm)-\tmp{\rm K}^2(\tm)\right] 
	\\
&+&2\tkappa^2\tm\left\{\left[{\rm E}(\tm)-{\rm K}(\tm)\right]^2-\tm{\rm K}^2(\tm)\right\}.
\end{eqnarray*}

The obtained system (\ref{sn-pot-ad1}) can be analyzed in the limiting cases. 
We start with the limit $\tilde{m}\to 0$ at $U_0<0$, by imposing the initial conditions 
\begin{eqnarray}
\label{init_cond_sn}
\tm(0)=0, \qquad \tkappa(0)=\kappa=2\pi/L
\end{eqnarray}
(the case becomes trivial in absence of the periodic potential, when $U_0=0$). Using  asymptotics (\ref{int-asymp-m}) and condition $\tm\ll 1$ we obtain from (\ref{sn-lat-L}) 
\begin{eqnarray}
	\label{link}
	\frac{\tkappa}{\kappa}=1+\frac{\tm}{4}.
\end{eqnarray}
Next, requiring $2\tkappa^2\tm\ll |U_0|$, what for $\kappa\sim 1$ is equivalent to $|\tkappa-\kappa|\ll U_0$, we deduce from the first of equations (\ref{sn-pot-ad1})
\begin{eqnarray}
\label{lin_as_sn}
	\tkappa^2&\approx& \kappa^2
	\left(1+\frac{4|U_0|\zeta}{16\kappa^2+|U_0|}\right)
\end{eqnarray}
where we used the condition $4U_0^2\zeta\ll 1$, necessary for the smallness of $\tm$. 
%Thus with time $\tkappa$ grows and $\tm$ approaches the unity. 

In the opposite limit $\tm\to 1$ one can use the asymptotics (\ref{int-asymp}) to obtain
\begin{eqnarray}
	\label{nln_as_sn}
	\tkappa\sim e^{\zeta/2},\qquad \tm=1-e^{-L\tkappa/2}.
\end{eqnarray}
Now the potential amplitude does not affect the asymptotics in the leading order. This is expectable behavior because it corresponds to large particle densities, when two body interactions dominate the effect of the periodic potential.

In Fig.\ref{sn-pot} we show examples of numerical solutions of equations (\ref{sn-pot-ad1}) for three different situations with respect to $U_0$ and intermediate values of the wave parameters. In all the cases the wave is transformed in a sequence of dark solitons. The external potential affects the velocity of the process. 
 
\subsection{BEC with a negative scattering length: cn-wave}
\label{sec_cn_latt}

Now we consider deformation of the periodic cn-wave resulting in creation of a train of bright matter solitons in a BEC with the negative scattering length, $\tsigma=-1$, loaded in OL (\ref{Vlatt}). 
The cn-wave unaffected by FR is given by the second row of the Table~\ref{tabletwo}.
Thus we look for a solution in the form
\begin{eqnarray}
	\label{cn_wave_pot}
	v_{\cn}=\sqrt{\tkappa^2\tm-\frac{U_0}{2}}e^{i(\tkappa^2(2\tm-1)-U_0)t}\cn (\tkappa x, \tm)\,.
\end{eqnarray}
Notice that, as above, choice of the magnitude of the OL amplitude has the constrain $U_0<2\tkappa^2 \tm$. 
Unlike in the previous case, now the value of the amplitude of the OL changes not only the initial atomic distribution but also the frequency of the solution. 

Now we have
\begin{eqnarray}
	\label{cn-latt-L}
	\begin{array}{l}
	\displaystyle{
	L=\frac{4{\rm K}(\tm)}{\tkappa}
	,}\\
	\displaystyle{
	N=\frac{4\tkappa^2\tm-2U_0}{\tkappa \tm}\left[{\rm E}(\tm)-\tmp{\rm K}(\tm)\right] 
	}
	\end{array}
\end{eqnarray}
and equations of the adiabatic approximation (\ref{wl}), (\ref{Neq}) can be written down in the form:
\begin{eqnarray}
	\label{cn-adiab-eqn}
\begin{array}{l}\displaystyle{
\frac{d\tkappa}{d\zeta}=\frac{\tkappa ( 2\tkappa^2\tm -U_0)}{\Delta_{\cn}}\left[{\rm E}(\tm) - \tmp{\rm K}(\tm)\right]^2,}\\
\displaystyle{
\frac{d\tm}{d\zeta}=\frac{2( 2\tkappa^2\tm-U_0)\tm\tmp}{\Delta_{\cn}}
%}\cr\displaystyle{\qquad \qquad\times
{\rm K}(\tm)\left[{\rm E}(\tm) -\tmp{\rm K}(\tm)\right]}
\end{array}
\end{eqnarray}
where
\begin{eqnarray*}
\Delta_{\cn}&=&U_0\left[{\rm E}^2(\tm)-\tmp{\rm K}^2(\tm)\right] 
	\\
&+&2\tkappa^2\tm\left\{{\rm E}^2(\tm)+\tmp {\rm K}(\tm) \left[{\rm K}(\tm)-2{\rm E}(\tm)\right]\right\}.
\end{eqnarray*}
 
As before we consider the limiting cases starting with the initial conditions (\ref{init_cond_sn}) what is possible only for negative $U_0$. The link (\ref{link}) holds in the case at hand, while the dynamics of $\tkappa$ is governed by the formula
\begin{eqnarray}
\label{lin_as_cn}
	\tkappa^2&\approx& \kappa^2
	\left(1+\frac{4|U_0|\zeta}{16\kappa^2-|U_0|}\right).
\end{eqnarray}
The obtained result reveals an interesting feature: depending on the amplitude of the potential barrier between two adjacent minima, $U_0$, the amplitude of the periodic wave is either increasing, if $-16\kappa^2<U_0 <0$, or decreasing, if $U_0<-16\kappa^2$. 
This is a general property, rather than the phenomenon observed in the limit of small $\tm$. 
Indeed, for a given $U_0$, the denominator $\Delta_{\cn}$ in (\ref{cn-adiab-eqn}) acquires zero value for some $\tilde{m}$ and $\tilde{\kappa}$, while the numerators in (\ref{cn-adiab-eqn}) are positive. In small vicinities of each of the curves, where $\Delta_{\cn}$ is close to zero the adiabatic approximation in the form (\ref{cn-adiab-eqn}) is not valid. 

To explain the described behavior we notice that decreasing of the wave amplitude is possible only for negative $U_0$. 
In the case under consideration, where $U_0<0$, atoms are collected around maxima of the potential (what happens due to attractive interactions and is impossible in the linear limit). Thus the condensate undergoes the effect of two forces acting in opposite directions: one originated by attractive interaction among particles and another force, tempting to separate condensates, due to the lattice potential. 
When the height of the potential barrier becomes larger than some critical value the second force becomes dominant, and the condensate tends to ``split'' in the vicinity of each maxima resulting in decrease of the amplitude and shift of the atomic density toward the potential minima (notice, that this situation reminds the case depicted in Fig.\ref{sn-pot}a).
 
Returning to the asymptotic (\ref{lin_as_cn}) we observe that $\tm$ becomes negative if $U_0<-16\kappa^2$ and thus the atomic density is approximated by the function
\begin{eqnarray}
	\label{cd_wave_pot}
	v_{\cn}=\sqrt{\frac{|U_0|}{2}-\tkappa^2|\tm|}e^{i(-\tkappa^2(2|\tm|+1)+|U_0|)t}
	\nonumber \\
	\times\,\mbox{cd} \left(\sqrt{1+|\tm|}\tkappa x, \frac{|\tm|}{1+|\tm|}\right)\,.
\end{eqnarray} 
 
\begin{figure}[ht]
%\vspace{0.5 true cm}
\epsfig{file=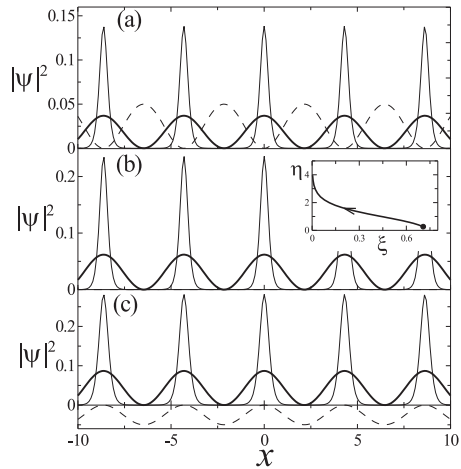,width=6cm}%=\columnwidth}
%\vspace{0.5 true cm}
\caption{Examples of cn-wave solutions at $\zeta=0$ (solid thick lines) and $\zeta=\zeta_{fin}$ (solid thin lines) in the periodic potential (dashed lines) for (a) $U_0=0.05$, and $\zeta_{fin}=4.5$; (b) $U_0=0$, $\zeta_{fin}=4$, and (c) $U_0=-0.05$, $\zeta_{fin}=3.5$. 
The inset in (b) shows dynamics of the parameters $\eta$ and $\xi$ given by (\ref{cn-adiab-eqn}). 
In all cases $m=0.11$ and $\kappa=0.75$.}
\label{cn-pot} 
\end{figure}

When the amplitude of the periodic wave growth, i.e. when $-16\kappa^2<U_0 <2\kappa^2m$, one observes generation of the train of bright solitons. 
An example of numerical solutions of equations of the adiabatic approximation for different intensities of periodic potential is presented in Fig.\ref{cn-pot}.

\subsection{BEC with a negative scattering length. dn-wave}
\label{sec_dn_latt}

As the last example of the adiabatic dynamics we consider evolution of the periodic dn-wave. 
As in the previous case the attractive nature of the interaction ($\tsigma=-1$) and application of the FR result in creation of a train of bright matter solitons. 
A solution of (\ref{modNLS1}) is searched now in the form (see last row of Table~\ref{tabletwo}) 
\begin{eqnarray}
	\label{adiabat_dn}
	v_{\dn}=\sqrt{\left(\tkappa^2-\frac{U_0}{2\tm}\right)}
	e^{i(\tkappa^2(2-\tm)-U_0/\tm)t}\dn(\tkappa x, m)
\end{eqnarray}
for which
\begin{eqnarray}
	\label{dn-latt-L}
	L=\frac{2{\rm K}(\tm)}{\tkappa},\quad
	N=\frac{2(2\tkappa^2\tm-U_0){\rm E}(\tm)}{\tkappa \tm}
\end{eqnarray}
and equations of the adiabatic approximation read:
\begin{eqnarray}
\label{dn-adiab-eqn}
\begin{array}{l}\displaystyle{
\frac{d\tkappa}{d\zeta}=\frac{\tkappa ( 2\tkappa^2\tm-U_0)}{\Delta_{\dn}}{\rm E}(\tm)\left[{\rm E}(\tm) -\tmp{\rm K}(\tm)\right],}\\
\displaystyle{
\frac{d\tm}{d\zeta}=\frac{2(2\tkappa^2\tm-U_0)\tm\tmp}{\Delta_{\dn}}{\rm E}(\tm){\rm K}(\tm)}
\end{array}
\end{eqnarray}
where
\begin{eqnarray*}
	\Delta_{\dn}&=&U_0\left[{\rm E}^2(\tm) + \tmp{\rm K}^2(\tm)\right]\\
&+&2\tkappa^2\tm\left[{\rm E}^2(\tm)-\tmp{\rm K}^2(\tm)\right].
\end{eqnarray*}

\begin{figure}[ht]
%\vspace{0.5 true cm}
\epsfig{file=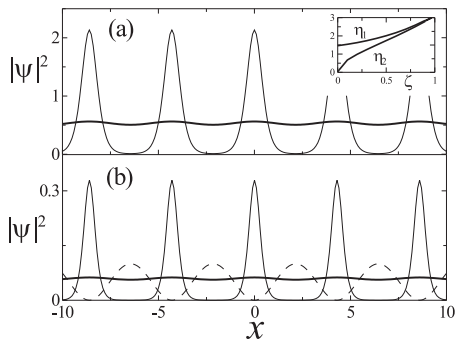,width=6cm}
%\vspace{0.5 true cm}
\caption{Examples of dn-wave solutions at $\zeta=0$ (solid thick lines) and $\zeta=\zeta_{fin}$ (solid thin lines) in the periodic potential (dashed lines) for (a) $U_0=0$, and $\zeta_{fin}=0.25$; (b) $U_0=0.1$, $\zeta_{fin}=3$. 
The inset in (a) shows the respective dynamics of parameters $\eta_{1,2}$ given by (\ref{dn-syst}). 
In all cases $m=0.11$ and $\kappa=0.75$.}
\label{dn-pot} 
\end{figure}

An example of dynamics of the dn-wave, obtained by numerical solution of the system (\ref{dn-adiab-eqn}) is shown in Fig.\ref{dn-pot}. 

\section{BEC in an OL subject to FR. Including boundary effects}
\label{num1D}

In the last two sections we considered situations, where the parabolic trap is long enough in the axial direction and dynamics of the BEC can be treated within the framework of the adiabatic theory. 
In other two cases, where boundary effects are relevant [Eqs. (\ref{nls1dcase2}) and (\ref{nls1_case4})] numerical analysis has to be employed. 
This rises the problem of a choice of adequate initial conditions (i.e. the initial conditions which in the absence of FR are stationary solutions, such that the whole dynamics is originated by the FR only). 
Case~2, i.e. the NLS equation with an effective mass (\ref{nls1dcase2}), was studied in \cite{prl90AKKB}. In the present section we are concerned with the Case~4. 
 
To this end, we impose the following temporal behavior of the nonlinearity \cite{bright}
\begin{equation}
g(t)=\exp(t/\tau)
\label{g_t}
\end{equation} 
where $\tau$ is a constant. Then wave evolution is governed by the equation [c.f. (\ref{nls1_case4})]
\begin{eqnarray}
i\psi_t=-\psi_{xx} + U_0 \sn^2(\kappa x,m)\psi 
+2\tsigma e^{t/\tau} |\psi|^2\psi
\nonumber \\
+\nu^2x^2\psi.
\label{Braz_nls3_sn} 
\end{eqnarray}

We also suppose that initial scattering length, $a_s(0)$ is small enough, such that the two-body interactions are negligible. This implies smallness of the renormalized density, $|\psi|^2\ll 1$. The aspect ratio is considered to be $\nu \sim 10^{-3}\div 10^{-4}$ what allows us to consider initial distributions in a form of modulated waves: $\psi(x,t=0)=\exp(-\nu x^2/2)u_{{\rm pq}}(x)$, close to the ground state in the absence of FR.

\subsection{Deformation of a cn-wave}
 
\begin{figure}[ht]
\epsfig{file=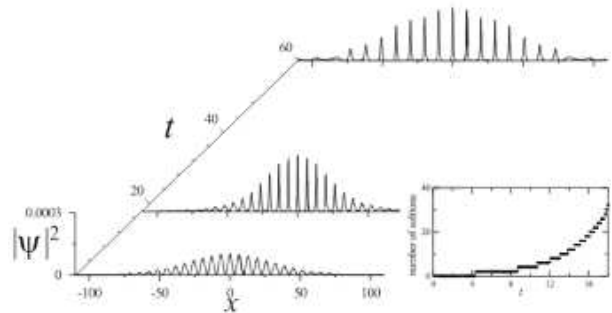,width=8cm}
\caption{ Dynamics of the cn-wave. 
%(a) thick and thin lines show initial ($t=0$) and intermediate ($t=19$) 
%profiles of the wave, respectively, and in (b) the potential is shown. 
The parameters are $\nu=0.0005$, $\tau=2$, $\kappa=0.5$, $m=0.1$, and $U_0=0.05$.
In the inset growth of the number of solitons is shown. 
\label{figcnv}} % Give a unique label
\end{figure}

First we consider deformation of a periodic cn-wave resulting in creation of a train of bright solitons in a BEC with negative scattering length embedded in OL (\ref{Vlatt}). 
A typical example of numerical solution of Eq.(\ref{Braz_nls3_sn}) with initial condition taken in a form of modulated cn-wave is presented in Fig.\ref{figcnv}. 
We show two subsequent regimes. 
First, on the time interval $0\leq t\leq 19$ the FR was switched on and after that at 
$19< t\leq 60$ all potentials and FR were switched off, and the condensate was allowed to expand (freely in one dimension).

%\vspace{0.5 true cm}

In Fig.\ref{figcnv} one can see that growth of the scattering length 
leads to sharpening of peaks of the distribution and to increase of their amplitudes. 
At an intermediate moment of time (here $t=19$) we obtain strongly pronounced train of pulses, which is naturally to associate with bright solitons.
Strictly speaking the concept of a soliton is mathematically well defined in integrable systems, like NLS equation (\ref{nls1D}), where each solution is associated with an eigenvalue of discrete spectrum of the ZS spectral problem (see Appendix A). 
In order to make use of this mathematical definition in our case, we reformulate it in the following way. 

Let us assume that a wave evolves during the time interval $0<t<t_{0}$. 
Then we assume that at $t_0$ the trap and lattice potentials are switched off and the output pulse, i.e. one at $t=t_0$, is governed by the unperturbed NLS equation.
Then we compute the number of NLS solitons generated by the wave profile given at $t=t_0$. 
To this end we substitute $\psi(x, t=t_0)$ into the ZS spectral problem and study its spectrum.

The result of this approach is shown in the inset of Fig.\ref{figcnv}.  
At each time step ($\Delta t=0.005$) in the interval $0<t\leq 19$ the function $\psi(x,t)$ is considered as an initial condition for the respective NLS equation. 
One can see that the initial cn-wave does not have solitons while increase of the nonlinear coefficient with time results in exponential growth of the number of solitons (to compute the number of solitons we used the numerical approach due to \cite{osborn}). 
%
%\begin{figure}[ht]
%\epsfig{file=cnv_3d.eps,width=7.5cm}%\columnwidth}
%\epsfig{file=cnv_num.eps,width=6cm}
%\caption{Dynamics of the number of solitons from the Fig.\ref{figcnv}. On each time step ($\Delta t=0.005$) in the interval $0<t\leq %19$ the function $\psi(x,t)$ considered as an initial condition $\psi_n$ for the respective NLS equation.
%\label{figcnv_num}} % Give a unique label
%\end{figure}

In order to visualize the generated solitons in Fig.\ref{figcnv} we show how the localized pulses, created due to the FR, propagate after all potentials are switched off at $t=19$. 
In Fig.\ref{figcnv} we can see that indeed the condensate consists of a set of solitary pulses propagating outwards the center without significant changes of their shapes. 
This behavior supports the fact that we indeed generated train of bright solitons. 
Meantime it should be mentioned that at $t=19$ the number of solitons calculated by solving spectral problem exceeds the number of solitons visible in Fig.\ref{figcnv}. 
This can be explained by the facts that peaks in Fig.\ref{figcnv} can present bound states consisting of several solitons which move with the same velocities and that small amplitude solitons are invisible on the scale of picture.

\subsection{Deformation of a dn-wave}

\begin{figure}[th]
\epsfig{file=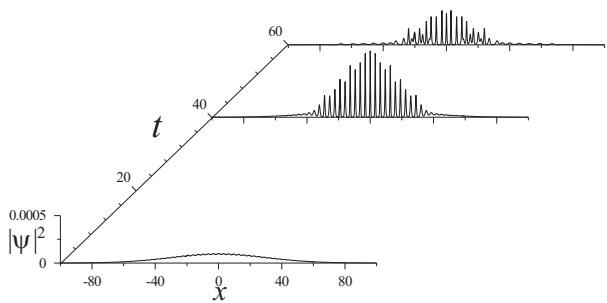,width=8cm}
\caption{ Dynamics of the dn-wave. 
%In the inset (a) thick and thin lines show initial ($t=0$) and intermediate ($t=40$) profiles of the wave which are affected by FR and in (b) the potential is shown. 
The parameters are $\nu=0.0005$, $\tau=4$, $\kappa=1$, $m=0.1$, and $U_0=0.2$. 
\label{figdnv}} % Give a unique label
\end{figure}

Now we consider deformation of the dn-wave resulting in creation of a train of bright 
matter solitons in a BEC with the negative scattering length. We solve Eq. (\ref{Braz_nls3_sn}) with initial condition taken as a dn-wave (third row in Table~III) modulated by the Gaussian background. 
As in the previous case, increase of the strength of the inter-particle interactions by means of FR, results in 
growth and compression of the distribution picks, transforming them in a train of bright solitons (see the insert in Fig.\ref{figdnv}). 

In contrast to the previous case, however, the generated train of solitons is not robust, and is destroyed after some time of free propagation, corresponding in Fig.\ref{figdnv} to time interval $t>t_0$ where the wave propagates freely. 
The difference in the behavior of trains of solitons can be understood if one takes into account that the dn-wave has a period of the lattice potential while cn-wave has a double period. 
As a result all bright solitons generated from the dn-wave are in-phase, while solitons generated from cn-wave have alternating phases (and thus the nearest neighbors are out-of-phase). 

\subsection{Deformation of an sn-wave}
\label{sec:snv}

As the last example of this section we consider the generation of dark solitons in a BECs with a positive scattering length and with initial condition taken in a form of the sn-wave (the first row of Table~III), modulated by the Gaussian background.
This case is very different from the two previous ones. 
The main reason is that now we are dealing with a periodic wave transforming in a train of dark solitons, which exist against a background. Either by change of the nonlinearity or by allowing free expansion of the condensate the background is strongly affected, what becomes an obstacle for numerical (and experimental) realization of the scenario described in the subsection~\ref{subsec:sn-period}. It turns out however that it is possible to reduce the effect of the condensate expansion. 

To this end we first consider the NLS equation with constant nonlinear coefficient ($g=1$) 
\begin{equation}
\label{app_nls3} i\partial_t \psi =- \partial^2_x \psi + \nu^2x^2\psi +2 |\psi|^2\psi,
\end{equation}
and consider stationary solution $\psi\approx u_{TF}(x)\, e^{-i \mu t}$ in the Thomas-Fermi approximation: $|u_{TF}|^2= (\mu-\nu^2 x^2)/2.$ 

Next we assume that FR is switched on and try to compensate growth of the nonlinearity by time dependence of the chemical potential $\mu(t)$. Then the respective solution is approximated by $\tilde\psi=\tilde{u}_{TF}(x)\, e^{-i {\cal M}(t) }$ where ${\cal M}(t)$ is to be determined. 
In order to compensate the change of the nonlinear part we also vary in time the strength of the parabolic potential $\nu\to \tilde\nu(t)$. Then
 the Thomas-Fermi approximation takes the form
$ |\tilde{u}_{TF}|^2=\frac{{\cal M}_t-\tilde\nu^2(t) x^2}{2g(t)}$.
In order to preserve the background profile during action of FR we require $u_{TF}$ and $\tilde{u}_{TF}$ to be equal what leads to the conclusion that the parameter $\tilde\nu(t)$, defining change of the aspect ration of the cigar-shaped trap, must change in time according to the law $\tilde\nu(t)=\nu \sqrt{g(t)}$.

Thus, ``stable'' generation of a train of dark solitons can be achieved by simultaneous application of the FR and the respective trap potential what is described by the model 
\begin{eqnarray}
i\partial_t\psi=-\partial^2_x\psi + U_0 \sn^2(\kappa x,m)\psi 
+ e^{t/\tau}\nu^2x^2\psi 
\nonumber \\
+2e^{t/\tau} |\psi|^2\psi.
\label{Braz_nls3_sn_1} 
\end{eqnarray}
 
\begin{figure}[ht]
\epsfig{file=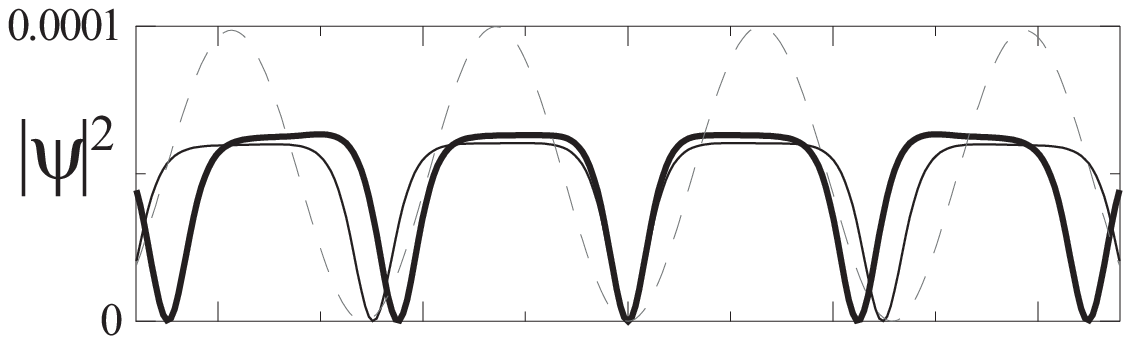,width=6cm} 
\caption{% Initial $t=0$ (thin line) and final $t=21$ (thick line) profiles of the condensate (a) and potential (b). 
Dynamics of dark solitons in the parabolic potential after switching off the FR and periodic potential at $t=21$. Dashed, thin and thick lines correspond to $t=0; 21; 25$. Parameters are $\nu=0.0002$, $\tau=2$, $\kappa=0.5$, $m=0.1$, and $U_0=0.05$.
\label{figsnv}}
\end{figure}

Examples of initial and final shapes of the condensate density affected by FR and adiabatically varying trap potential within the framework of the model (\ref{Braz_nls3_sn_1}) are shown in Fig.\ref{figsnv}.

To identify a train of dark solitons, avoiding free expansion of the condensate, we switch off OL, FR and modulation of the trap potential, holding the last one unchanged. 
As one can see form the Fig.\ref{figsnv}, where the switching time is $t=21$, the dark solitons start to move toward the center of the condensate. 
This happens because of reflection of solitons by the potential boundaries \cite{BK03}, which due to introduced modulation of $\nu$ the trap potential has become more narrow than it was at $t=0$. 
The difference in the depth of each of solitons causes different velocities and distance which solitons pass, what is also  seen in Fig.\ref{figsnv}.

\section{3D numerical simulations. The general case}
\label{3D}

As it is clear 1D models considered above cannot account all features of the BEC dynamics, what becomes especially relevant in the case of growing nonlinearity. In particular, the later can change significantly the background resulting in failure of the low density approach. 
That is why we study numerically 3D GP equation (\ref{GPE}) for a cigar-shaped condensate having radially symmetric configuration. 
Using the same dimensionless variables as in Eq.(\ref{GPE1}) we rewrite the GP equation in the form 
\begin{eqnarray}
i\partial_t \Psi&=&-\left(\partial^2_x +\frac{1}{r_\bot}\partial_{r_\bot} r_\bot\partial_{r_\bot}\right)\Psi+(r_\bot^2+\nu^2 x^2)\Psi\nonumber\\
&+& U_0\kappa^2 \sn^2(\kappa x, m)\Psi +8\pi\sigma e^{t/\tau} |\Psi|^2\Psi\, \label{3DGP_dim}
\end{eqnarray}
where all tildes are suppressed and it is assumed that the nonlinearity varies according to (\ref{g_t}).

By using the assumption about initially weak nonlinearity  one can consider initial profiles of the condensate as 1D periodic solutions $u_{\rm pq}(x)$, taken from the Table~III, where $U_0$ must be substituted by $\kappa^2 U_0$ due to the normalization accepted in Sec. \ref{ini} [see Eq.(\ref{Vlatt})], modulated by the 3D Gaussian envelope:
\begin{eqnarray}
|\Psi(x,{\bf r}_\bot,t=0)|^2=u_{\rm pq}^2(x) e^{-r_\bot^2-\nu x^2}. \label{3D_ini}
\end{eqnarray}
Such a distribution is close to the ground state and its temporal evolution occurs mainly due to change of the scattering length.

Like in 1D case we use implicit Crank-Nicolson scheme for numerical integration of the radially symmetric 3D GP equation. The scheme stability are fulfilled by controlling the ratio between time and space steps and the results are confirmed by using different time and space grids. On each time step of calculations in each site of the grid  we check the convergence of the numerical scheme. In all 3D calculation we used grid of 400$\times$400 points with the spatial step~0.2 and temporal step 0.001.

\subsection{Dynamics of cn- and dn-waves in a BEC with a negative scattering length}

\begin{figure}[ht]
\epsfig{file=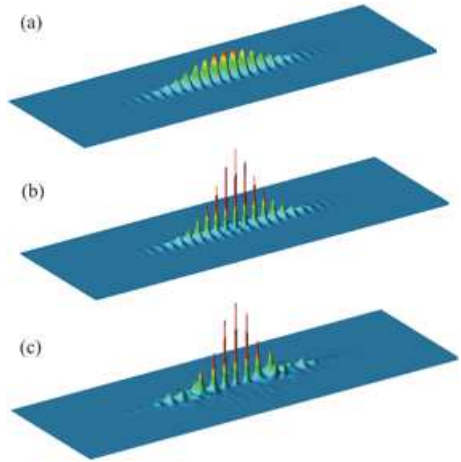,width=6cm} 
\caption{Dynamics of the condensate density, $|\Psi({\bf r})|^2$, obtained by numerical solution of GP equation (\ref{3DGP_dim}) with attractive interactions ($\sigma=-1$). 
Initial density is taken as (\ref{3D_ini}) where parameters of cn-wave are $m=0.1$, $\kappa=0.5$ and $\nu=0.02$. The amplitude of OL is $U_0=0.2 E_R$ with $E_R\approx 0.7\,$peV.
(a) Initial, (b) intermediate ($t_0=4.4$~ms) and (c) final ($t_f=9$~ms) density profiles.  The characteristic time of FR $\tau= 1$~ms. 
The size of each image is $180\times 40\, \mu$m. %The value of the distribution picks is shown in Fig.\ref{figDcnv}.
\label{figDDDcnv}}
\end{figure}

First we consider adiabatic dynamics of a BEC with initial density distribution (\ref{3D_ini}) where $u_{\rm pq}(x)$ is taken to be the cn-wave (see Table~III).
To be specific, the numerical results reported below in the physical units correspond to ${\cal N}\approx 5\times 10^{3}$ atoms of $^{7}$Li loaded in an elongated trap with $\omega_0=2\pi\times 1.2\, $Hz and $\omega_\bot=2\pi\times 57.5\, $Hz 
(the aspect ratio being $\nu\approx 0.02$). Taking into account that 
by changing the magnetic field near resonant point 725~G one can change scattering length by order $10^2\div 10^3$ \cite{bright}, the initial magnitude of the scattering length is chosen to be $a_s(0)=-0.1\, $nm. 
This corresponds to transfer of the condensate from quasi-linear limit to the limit of relatively strong two-body interactions.
The respective initial density distribution is shown in Fig.\ref{figDDDcnv}a.

While FR is switched on (i.e. for $t<4.4$~ms) the periodic density distribution displays sharpening in each potential well acquiring a shape of quasi-1D soliton train at $t=4.4$~ms (Fig.\ref{figDDDcnv}b). % and Figs.\ref{figDcnv}a,b. 
As one can see from Figs.\ref{figDDDcnv}a,b, 
%Figs.\ref{figDcnv}a,b, 
appearing of localized excitations is accompanied by appreciable narrowing of the transverse profile of the condensate (evidently not taken into account in the 1D approximation developed above). 
Change of the axial dimension of the condensate is negligible. 
We also notice that due to choice of relatively weak initial nonlinearity collapse does not occur. 

Next, at the intermediate time $t_0=4.4$~ms [what corresponds to $a_s(t_0)\approx 81.5 a_s(0)$] we switch off FR, OL and trap potential in the longitudinal direction (leaving only parabolic potential in the radial direction). 
This allows condensate to expand freely in the axial direction (a kind of a waveguide configuration).  A density distribution after 4.6~ms of free unidirectional expansion is shown in Fig.\ref{figDDDcnv}c. 
Now the transverse profile of the condensate changes weakly 
while localized pulses move outwards the center of the condensate. 
In the 1D approximation such pulses were identified as solitons. 
Now we also observe that low amplitude solitary pulses (they are located on the wings of the condensate) dispersion dominates nonlinearity and one observes spreading out of each pulse. Large amplitude pulses preserve well pronounced spatially localized structure, while moving along the effectively created waveguide in the condensate. 

\begin{figure}[ht]
\epsfig{file=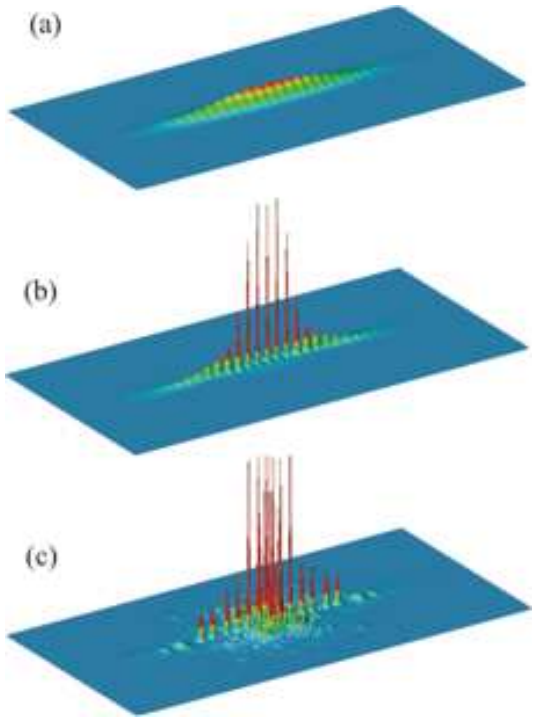,width=7cm} 
\vspace{1cm}
\caption{Dynamics of the condensate density, $|\Psi({\bf r})|^2$, obtained by numerical solution of GP equation (\ref{3DGP_dim}) with attractive interactions ($\sigma=-1$). The initial density distribution (\ref{3D_ini}) is taken as a dn-wave with $m=0.1$, $\kappa=1$ and with $\nu=0.02$. 
(a) Initial, (b) intermediate ($t_0=5.2$~ms) and (c) final ($t_f=10$~ms) density profiles of the condensate.
The amplitude of OL is $U_0=0.2 E_R$ with $E_R\approx 0.35\,$peV, and 
And $\tau=2$~ms. The size of each image is $180\times 80\, \mu$m.
\label{figDDDdnv}}
\end{figure}
 
Next we consider evolution of the dn-wave (where in the initial density distribution (\ref{3D_ini}) $u_{\rm pq}(x)\equiv u_{\dn}(x)$ from Table~III). 
Now we use the same magnetic trap as in the previous case, taking ${\cal N}\approx 10^4$ of $^7$Li atoms. 
Numerical simulations are carried out in the same way as this was done in the previous example.
FR is switched off at time $t_0=5.2$~ms (what corresponds to $a_s(t_0)\approx 14 a_s(0)$) and free 1D expansion of the condensate along the waveguide during 4.8ms between $t_0$ and $t_f=10$~ms is considered.

While the FR is switched on one observes strong sharpening of the distribution maxima (much more pronounced than in the case of the  cn-wave shown in Fig.\ref{figDDDcnv}) with significant narrowing of the transverse profile (see Fig.\ref{figDDDdnv}b).
Meantime, during the free expansion, the condensate displays less regular structure of the train of solitary pulses (see Fig.\ref{figDDDdnv}c),
what was predicted by the 1D theory and explained by the fact that now we have a sequence of the in-phase neighbor pulses. Another property, found numerically is that oscillatory behavior of the transverse distribution appears (the effect neglected by 1D theory).

\subsection{Dynamics of an sn-wave in a BEC with a positive scattering length}

\begin{figure}[ht]
\epsfig{file=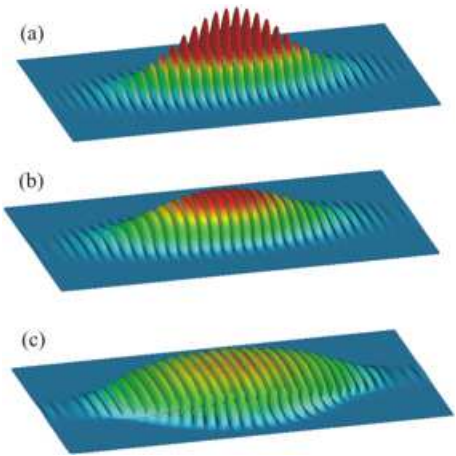,width=6cm} 
\caption{Dynamics of the condensate density obtained by numerical solution of GP equation (\ref{3DGP_dim}) with repulsive interactions ($\sigma=1$). 
Initial density is taken as the sn-wave with $m=0.1$ and $\kappa=0.75$ modulated by the Gaussian function with $\nu=0.1$. 
(a) Initial, (b) intermediate ($t_0=30$~ms) and (c) final ($t_f=45$~ms) density profiles of the condensate. 
The OL amplitude is $U_0\approx -0.2 E_R$ with $E_R\approx 0.034\,$peV, and $\tau\approx 6$~ms. 
The size of each image is $150\times 60\, \mu$m.
\label{figDDDsnv}}
\end{figure}

As a final example we consider the case of a BEC with repulsive interactions ($\sigma=1$) where in the initial density distribution (\ref{3D_ini}) the solution $u_{\rm pq}$ is taken in a form of the sn-wave (i.e. as $u_{\sn}$ from Table~III). 
In Fig.\ref{figDDDsnv} 
we present density profiles of the condensate of ${\cal N}\approx 5\times 10^4$ atoms of $^{87}$Rb, confined by the trap with axial $\omega_0=2\pi\times 0.11\, $Hz and radial $\omega_\bot=2\pi\times 1.17\, $Hz harmonic oscillator frequencies, at different moments of time. It is supposed that initially two body interactions are negligible. 
The last condition for the given number of atoms can be achieved by proper choice of the external magnetic field. For example, according to \cite{FR} in the case of $^{87}$Rb atoms the magnetic field about 170~G 
results in $a_s(0)=0.1$~nm, what according to (\ref{small_pa}) gives the small parameter $\epsilon\approx 0.38$, providing the required weakly nonlinear limit. Then by approaching the resonant point ($\approx 156~G$) one can change magnitude of scattering length by $10^3$ times.

As before FR is switched on at $t=0$ leading to the evolution shown for $t_0=30$~ms in Figs.\ref{figDDDsnv}a,b. One observes decreasing of the amplitude of the density and formation of a sequence of well pronounced excitations having the shape of a train of quasi-one-dimensional dark solitons.

As before at some time, precisely at $t_0=30$~ms when the scattering length achieves the value $a_s(t_0)=148 a_s(0)$, we switch off both FR and OL, but unlike in the previous cases, we leave both the transversal and the axial parabolic traps. 
The condensate continues to expand in the transverse direction what leads to decrease its density and strong deviation from the original quasi-one-dimensional geometry.

\subsection{3D dynamics with 1D configuration. Comparison}

\begin{figure}[ht] 
\epsfig{file=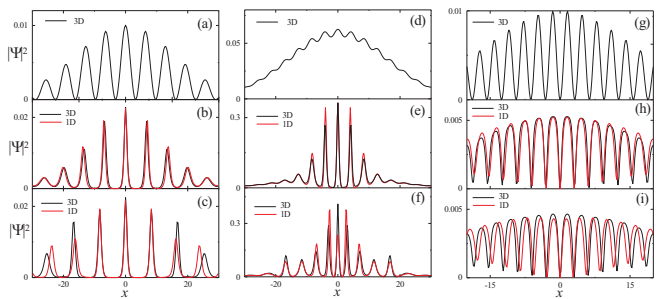,width=\columnwidth} 
\caption{ The cn-wave dynamics: (a) Initial, (b) intermediate ($t_0=22$) and (c) final ($t_f=42$) density profiles. The parameters are $U_0=0.03$, $m=0.1$, $\kappa=0.75$,  $\nu=0.001$, $\tau= 5$, and $\epsilon\approx 0.071$. The dn-wave dynamics: (d) Initial, (e) intermediate ($t_0=27.5$) and (f) final ($t_f=36$) density profiles. The parameters are $U_0=-0.1$, $m=0.1$, $\kappa=0.75$,  $\nu=0.002$,  $\tau= 10$, and $\epsilon\approx 0.068$. 
The sn-wave dynamics: (g) Initial, (h) intermediate ($t_0=7$) and (i) final ($t_f=9$) density profiles. The parameters are $U_0=-0.02$, $m=0.001$, $\kappa=1$, $\nu=0.002$,  $\tau= 1$, and $\epsilon\approx 0.1$. 
\label{fig_1d_3d_cnv}}
\end{figure}

Numerical study performed in the previous subsections used elongated traps with typical for experiments values of the linear oscillator frequencies.
While they illustrated qualitative agreement with prediction based on 1D models, they strictly speaking did not satisfy completely the conditions of the quasi-1D dynamics because of relatively wide transverse dimension. As a consequence 3D effects were not negligible especially at large nonlinearities.
It is therefore a relevant problem to provide quantitative comparison of 1D and 3D models. We will do this in details for the 4-th Case listed in Table~I, when the transverse size of the condensate is much smaller than the lattice constant. 
More specifically, we compare dynamics of the 3D condensate in the longitudinal direction governed by (\ref{3DGP_dim}) with dynamics of its 1D counterpart described by Eq.(\ref{GPE4}). We use the dimensionless variables of 3D problem, introduced in the last subsection, and take into account connections between 1D and 3D variables, coming from the multiple-scale expansion: $x_{1D}\leftrightarrow \epsilon x_{3D}$, 
 $t_{1D}\leftrightarrow \epsilon^2 t_{3D}$, $\Psi_{1D} \leftrightarrow \Psi_{3D}/\epsilon$, $\nu_{1D}\leftrightarrow \nu_{3D}/\epsilon^2$.
Naturally, for comparison we use the same initial distributions in the longitudinal directions taken in a form of cn- [Figs.\ref{fig_1d_3d_cnv}~(a)-(c)] and dn- waves [Figs.\ref{fig_1d_3d_cnv}~(d)-(f)] for a BEC with attractive interactions and in a form of an sn-wave [Figs.\ref{fig_1d_3d_cnv}~(g)-(i)] in the case of repulsive interactions.
 
In  Fig.\ref{fig_1d_3d_cnv} one observes a remarkable accuracy of the 1D approximation, small deviations from which occur due to the transverse dynamics of the condensates, which becomes more pronounced at large nonlinearities.  

\section{Conclusion}
\label{concl}
 
To conclude we investigated formation of trains of bright and dark matter solitons in an elongated Bose-Einstein condensates embedded in optical lattices by means of adiabatic change of the value of the scattering length. 
For a strongly elongated cigar-shaped condensate we considered reductions of the original Gross-Pitaevskii equation to different one-dimensional models, depending on the parameters of the problem (the results are summarized in Table~I). 
This was done in a controlled manner by using multiple-scale expansion method. 
Special attention has been paid to discussion of the link between periodic and solitary wave solutions of the one-dimensional nonlinear Schr\"odinger equation, what justified an idea how to affect a periodic matter wave loaded in a lattice, in order to create a train of solitons. 
In a number of cases, where boundary effects are negligible, the phenomenon can be described analytically using the adiabatic approximation. 
When both parabolic and periodic potentials are included explicitly in the one-dimensional model we had to involve numerical simulations in order to describe wave evolution. 
The consideration was concentrated on the three types of waves for which analytical expressions are available: cn-, dn- and sn-waves. 
Finally, the process of creation of the trains of bright and dark solitons by means of varying scattering length using Feshbach resonance in the presence of parabolic and periodic potentials was studied numerically withing the framework of the radially symmetric three-dimensional Gross-Pitaevskii equation. 
It has been shown that for the most typical experimental parameters one-dimensional approximation gives relatively good qualitative description of the dynamics. 
Meantime, it turned out that full three-dimensional consideration may be necessary because of significant deformation of the transverse structure of the condensate, especially in the case of dark solitons, which are very sensitive to the amplitude and structure of the background, affected by the Feshbach resonance. 

If all conditions of the applicability of the multiple scale expansion are satisfied, one observe remarkable quantitative agreement between the real three-dimensional models and it one-dimensional counterpart.

\section*{Acknowledgments}

V.V.K. is grateful to O. Morsch for sending the review paper \cite{MO} prior its publication. Work of V.A.B. has been supported by the FCT fellowship SFRH/BPD/5632/2001. 
 
\appendix

\section{Some elements of the Inverse Scattering Technique}
\label{IST}

The starting point of the IST is the fact that Eq.~(\ref{nls1}) appears as a compatibility condition for two systems of differential equations (see e.g. \cite{Faddeev})
\begin{eqnarray}
	\label{UV}
	|f\rangle_x={\bf U}|f\rangle,\qquad |f\rangle_t={\bf V}|f\rangle\,
\end{eqnarray}
where $|f\rangle=\mbox{col}(f_1,f_2)$, $f_{1,2}=f_{1,2}(x,t)$, 
\begin{eqnarray}
\label{U}
	{\bf U}&=&\left(
	\begin{array}{cc}
	\frac{\lambda}{2i} & \sqrt{\tsigma}\bar{\psi} \\
	\sqrt{\tsigma}\psi & -\frac{\lambda}{2i}
	\end{array}
	\right)\,, 
	\\
	\label{V}
	{\bf V}&=&\left(
	\begin{array}{cc}
	-\frac{\lambda^2}{2i} +i\tsigma |\psi|^2 & -\sqrt{\tsigma}(i\bar{\psi}_x+\lambda \bar{\psi})\\
	\sqrt{\tsigma}(i\psi_x -\lambda \psi)& \frac{\lambda^2}{2i}-i\tsigma|\psi|^2
	\end{array}
	\right)\,,
\end{eqnarray}
and $\lambda$ is called a spectral parameter. 
The first of systems (\ref{UV}) is referred to as the Zakharov-Shabat (ZS) spectral problem.
 
Let us introduce $\langle f|=(-f_2,f_1)$ and define the inner product~\cite{kawata} 
$	\langle f|g\rangle= f_1 g_2 - f_2 g_1 $
and a projector matrix
\begin{eqnarray}
	{\bf P}\equiv|f\rangle\langle g|=
	\left( \begin{array}{cc}
	-f_1g_2 & f_1g_1 \\ -f_2g_2 & f_2g_1 
	\end{array}\right).
\end{eqnarray}

If $|f\rangle$ and $|g\rangle$ are two independent solutions of each of the systems (\ref{UV}), then $\langle f|g\rangle$ is a Wronskian of each of them, and hence it does not depend neither on $x$ nor on $t$, i.e. $\langle f|g\rangle=p(\lambda)$, where $p(\lambda)$ is a function of the spectral parameter. 
Let $|f\rangle=\mbox{col}(f_1,f_2)$ be a solution of (\ref{UV}) with a given $\psi$. Then the symmetry properties of the matrices ${\bf U}$ and ${\bf V}$ imply that $|g\rangle=\mbox{col}(\bar{f}_2,\tsigma \bar{f}_1)$ is a solution of (\ref{UV}), as well. Using the pair of the solutions $|f\rangle$ and $|g\rangle$ one obtains that $\langle f| g\rangle=\tsigma|f_1|^2-|f_2|^2=p(\lambda)$.
 
Next, one ensures that ${\bf P}$ solves equations
\begin{eqnarray}
	\label{eq_proj}
	{\bf P}_x=[{\bf U},{\bf P}],\qquad {\bf P}_t =[{\bf V}, {\bf P}],
\end{eqnarray}
from which it follows that
\begin{eqnarray}
\label{p2}
	\Delta P^2-4 P_{12}P_{21}= P(\lambda)
\end{eqnarray}
where $\Delta P=P_{22}-P_{11}$, $P_{ij}$ being entries of the matrix ${\bf P}$, and $P(\lambda)$ is a function of $\lambda$ only.
For the chosen pair of solutions one verifies that $P_{21}=\tsigma \bar{P}_{12}$. 
Thus $P_{12}P_{21}=\tsigma |f_1|^2|f_2|^2$ and $\Delta P=\tsigma |f_1|^2+|f_2|^2$. 
Consequently, $P(\lambda)=p^2(\lambda)$ is a real positive function of $\lambda$. 

Noticing that ${\bf U}$ and ${\bf V}$ are matrix polynomials in $\lambda$, one can conjecture that the simplest nontrivial solution $|f\rangle$, and hence ${\bf P}$, should also be searched in forms of polynomials with respect to $\lambda$. 
 As far as ${\bf V}$ is a quadratic (matrix) polynomial of $\lambda$, one has to consider $|f\rangle$ to be a second (or higher) order polynomial with respect to $\lambda$. Since ${\bf P}$ is quadratic with respect to the elements of $|f\rangle$ one concludes that the simplest nontrivial form of $P(\lambda)$ must also be a polynomial of the forth degree of $\lambda$: 
\begin{eqnarray}
	\label{pl}
	P(\lambda)&=&(\lambda-\lambda_1)(\lambda-\lambda_2)
	(\lambda-\lambda_3)(\lambda-\lambda_4)
	\nonumber \\
	&=&\lambda^4+c_3\lambda^3+c_2\lambda^2+c_1\lambda+c_0.
\end{eqnarray}
Here $\lambda_{j}$($j=1,...,4$) are given complex numbers, and $c_j$ are trivially expressed through $\lambda_j$. Then, $\xi_j=\mbox{Re}\lambda_j$ and $\eta_j=\mbox{Im}\lambda_j$ can be considered as parameters characterizing respective solutions of the nonlinear problem (\ref{nls1}). The requirement for $P(\lambda)$ to be real imposes four constrains on the parameters $\lambda_j$. Indeed, considering $\bar{P}(\lambda)$ one concludes that either (i) all $\lambda_j$ are real, i.e. $\eta_j=0$ and $\xi_j$ parametrize the problem, or (ii) two of them are real and two ones are complex conjugated, say Im$\lambda_1$=Im$\lambda_2=0$ and $\lambda_3=\bar{\lambda}_4$, then the parameters are $\{\xi_1, \xi_2, \xi_3, \eta_3\}$, or (iii) there are two pairs of complex conjugated parameters, say $\lambda_1=\bar{\lambda}_2$ and $\lambda_3=\bar{\lambda}_4$, i.e. the parameters are $\{\xi_1, \xi_3, \eta_1,\eta_3\}$. 

Formulas (\ref{p2}) and (\ref{pl}) suggest that $P_{ij}$ should also be searched in a form of polynomials with respect to the parameter $\lambda$: 
\begin{eqnarray}
\label{assump}
\Delta P=\lambda^2+\lambda \Delta_1+\Delta_0,\qquad 
	P_{12}=\lambda P_{12}^{(1)}+P_{12}^{(0)}
\end{eqnarray}
where explicit form of the coefficients $\Delta_j$ and $P_{12}^{(j)}$ are to be found.
Comparing powers of $\lambda$ in (\ref{p2}) we obtain: 
\begin{eqnarray}
\label{connect1}
&&\Delta_1=\frac{c_3}{2},\quad \Delta_0=2\tsigma |P_{12}^{(1)}|^2+\frac 12 \left(c_2-\frac 14 c_3^2\right),
\\
 \label{connect1_2}
&&
 c_3\Delta_0-4\tsigma\left(\bar{P}_{12}^{(1)}P_{12}^{(0)}+\bar{P}_{12}^{(0)}P_{12}^{(1)}\right)=c_1,
\\
 \label{connect1_1}
&&\Delta_0^2-4\tsigma |P_{12}^{(0)}|^2=c_0. 
\end{eqnarray}
From the first of Eqs. (\ref{eq_proj}) we compute the equation for $\partial_x P_{12}$ whose consistency with representations (\ref{assump}) and (\ref{connect1}) results in the following connecting formulas
\begin{eqnarray}
	\label{connect2a}
	& & \bar{\psi}=\frac{i}{\sqrt{\tsigma}} P_{12}^{(1)}, \\
	\label{connect2b}
	& & \bar{\psi}_x=\frac{1}{\sqrt{\tsigma}}P_{12}^{(0)}+\frac{i}{2}c_3, \\
	\label{connect2c}
	& & \bar{\psi}_{xx}=\bar{\psi}\Delta_0 .
\end{eqnarray}
Comparing (\ref{connect2c}) with (\ref{nls1D}) one verifies that the two equations coincide subject to the conditions $ c_2=2\omega$ and $c_3=0$ what is the same as 
 \begin{eqnarray}
	\label{param_om}
	 \omega&=&\frac 12 \sum_{i,j=1 \atop i\neq j}^4 
	\lambda_i\lambda_j, 
\\		 
	\label{zero}
	0&=&\lambda_1+\lambda_2+\lambda_3+\lambda_4 .
\end{eqnarray}
%It worth pointing out here that in the case of more general solution (\ref{subs2}) the %sum of the parameters $\lambda_j$, i.e. coefficient $c_3$, determines the velocity %$V$, and thus (\ref{zero}) expresses the fact that the velocity of the solution we are %looking for is equal to zero.

It follows from (\ref{connect2a}) that $\bar{P}_{12}^{(1)}$ is a solution of (\ref{nls1}). 
Taking into account the particular form of solution (\ref{subs1}) one immediately concludes from (\ref{connect1_2}), (\ref{connect2a}), (\ref{connect2b}), and (\ref{zero}), that $c_1=0$. 
Finally, comparing (\ref{connect1_1}) with (\ref{solut1}) and taking into account (\ref{subs1}) one obtaines $C=\frac{\tsigma}{16}(c_2^2-4c_0)$ what is the same as 
\begin{eqnarray}
\label{param_c}
	C=\frac{\tsigma}{4}(\omega^2-\lambda_1\lambda_2\lambda_3\lambda_4 ).
\end{eqnarray}
The last condition leaves us with only two independent parameters. 

\begin{figure}[ht]
\vspace{1 true cm}
\center
\scalebox{0.3}[0.27]{\includegraphics{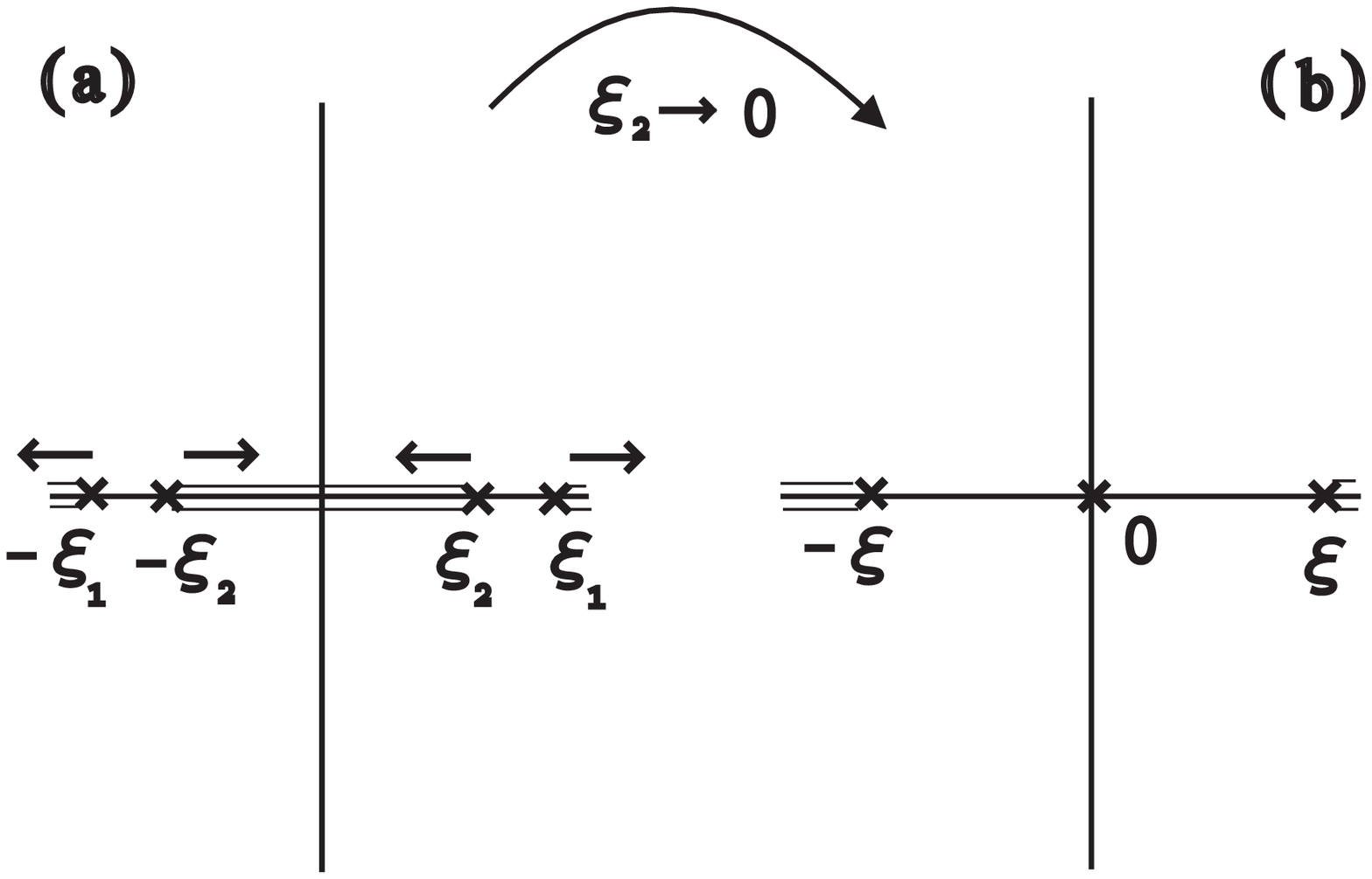}}\\
\scalebox{0.3}[0.27]{\includegraphics{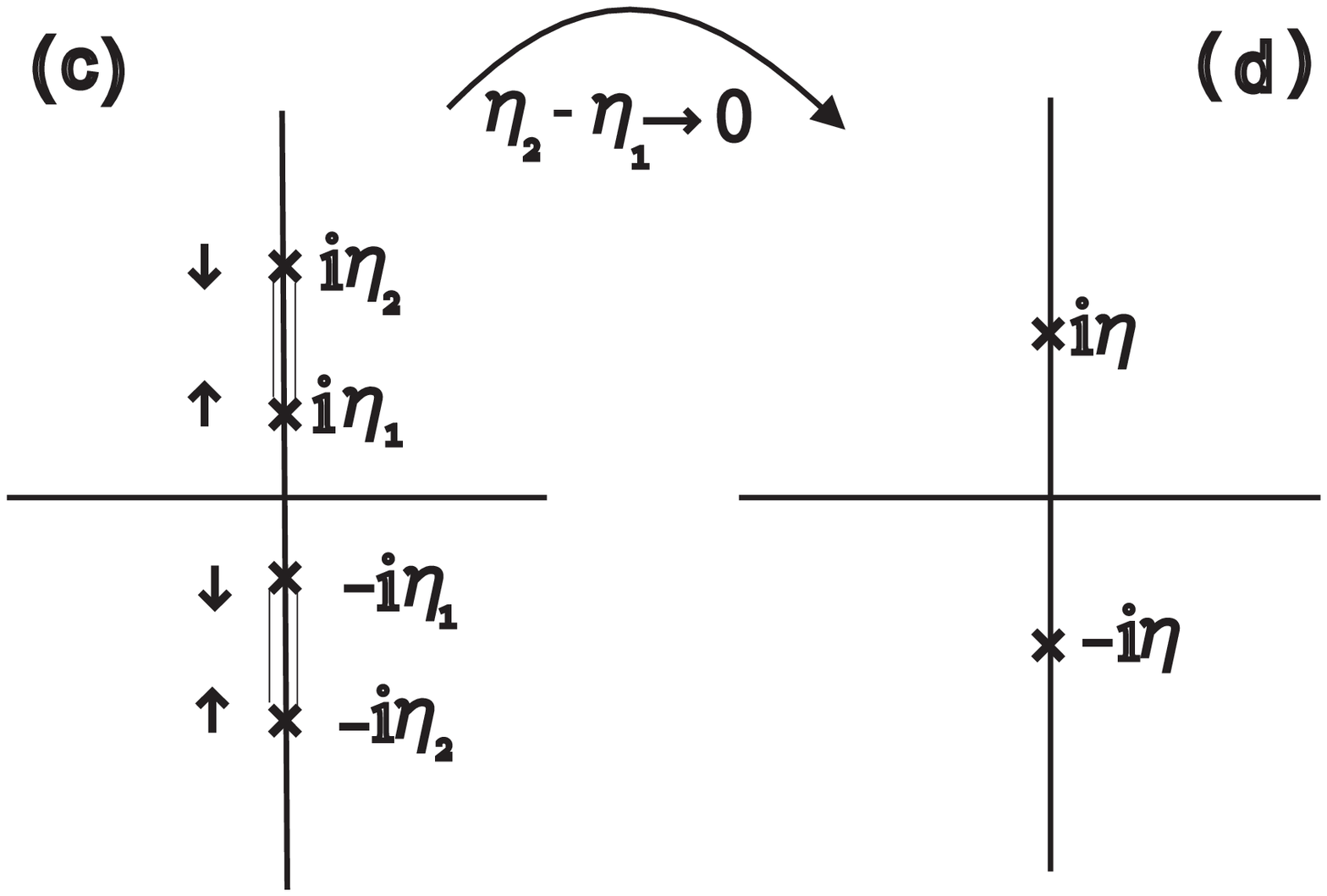}}\\
\scalebox{0.3}[0.27]{\includegraphics{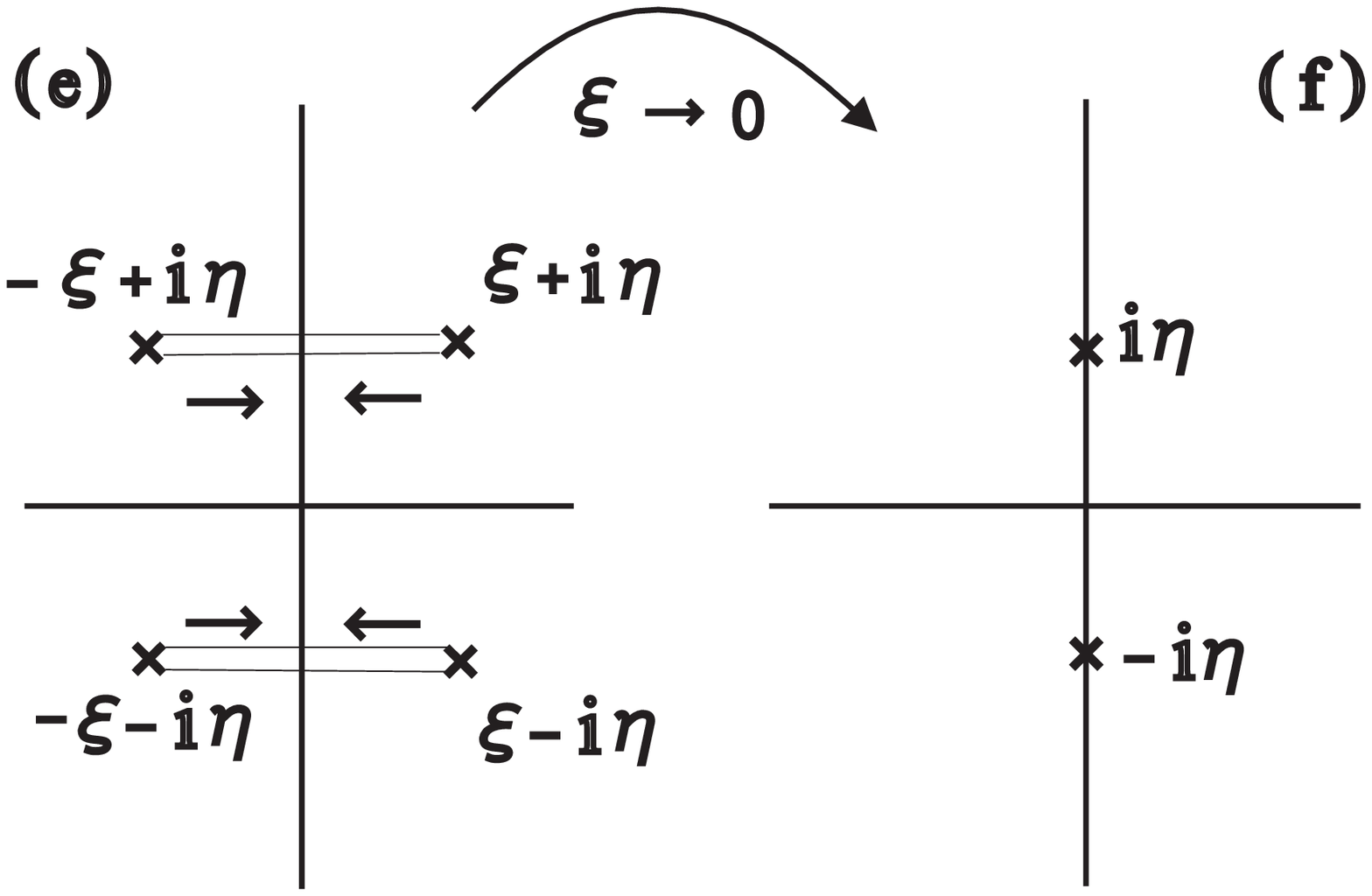}}\\
\caption{The Riemann surface of the function $p(\lambda)=\sqrt{P(\lambda)}$ consists of two complex plains, one of which is shown in Figs. (a) for sn-wave: $\lambda_1=-\lambda_3=\xi_1$, $\lambda_2=-\lambda_4=\xi_2$; (c) for dn-wave: $\lambda_1=\bar{\lambda}_3=i\eta_1$, $\lambda_2=\bar{\lambda}_4=i\eta_2$; and (e) for cn-wave: $\lambda_1=-\bar{\lambda}_3=-\lambda_2=-\bar{\lambda}_4=\xi+i\eta$. The right panels show the modification of the Riemann surface subject to the limiting transition $m\to 1$ shown in the last column of the Table~\ref{simplest}. The lines connecting the parameters $\lambda_j$ show cuts of the respective two-sheet Reamann surfaces.}
\label{figone} % Give a unique label
\end{figure}

A convenient representation of the results is given by the
locations of the parameters $\lambda_j$ on the Riemann surface of the function $p(\lambda)$, which is shown in Fig.\ref{figone} (Figs.\ref{figone} (a), (c), and (e) correspond to the above cases (i), (ii), and (iii)).

Let us now assume that by some means one can provide shifts of the spectral parameters in the complex plane. More specifically, we consider three situations corresponding to collapse of two pairs of the branching points, $\lambda_j$, as it is shown by the arrows in Fig.\ref{figone}. 

At $\tsigma=1$, subject to transition $\xi_2\to 0$ and $\xi_1\to\xi=2\rho$ (Fig.\ref{figone}b) the Riemann surface of $p(\lambda)$ is transformed in the Riemann surface of the spectral parameter of the ZS problem, subject to the boundary conditions  $\lim_{x\to\pm\infty}\psi=\rho e^{\pm i\varphi}$, where $\rho$ and $\varphi$ are real constants. Then the passage between (a) and (b) illustrates deformation of the branching points $\xi_{1,2}$ resulting in transformation of the sn-wave into a single dark soliton. The point $\xi_2=0$ corresponding to the discrete spectrum of the ZS spectral problem, which determines a static dark soliton, while points $\pm\xi_1= \pm 2\rho$ are transformed into edges of the continuum spectrum.

In the case of the NLS equation with $\tsigma=-1$ and zero boundary conditions, $\lim_{x\to\pm\infty}|\psi|=0 $, bright soliton solutions are determined by discrete spectrum located in the upper half-plane of the spectral parameter. 
If a soliton is static, the only discrete eigenvalue is placed on the imaginary axis, and is designated as $i\eta$ (Figs.\ref{figone}c and d). 
One can arrive at such situation by shifting the branching points of corresponding dn-wave as it is shown on the passage between Figs.\ref{figone}c and d. 
The points $\eta_{1,2}$ collapse into a point $\eta$ corresponding to the discrete eigenvalue of the ZS spectral problem. Alternatively, one can arrive at a static bright soliton solution, starting with a cn-wave, and providing displacements of the branching points as it is shown by the arrow between Figs.\ref{figone}e and f.

%\section{$UV$-pair}
%\label{app-UV}

\section{Some formulas from elliptic integrals}
\label{elliptic}

Throughout the text we use the asymptotic of the elliptic integrals (see e. g. \cite{Abramowitz,Lawden}) when $m\to 1$:
\begin{eqnarray}
	\label{int-asymp}
	\begin{array}{l}
	\displaystyle{	
	{\rm K}(m)=\ln\frac{4}{\sqrt{1-m}}\left[1+O(1-m)\right],
	}
	\\
	\displaystyle{	
	{\rm E} (m)=1+O\left[(1-m)\ln(1-m)\right],
	}
	\end{array}
\end{eqnarray}
and when $m\to 0$
\begin{eqnarray}
	\label{int-asymp-m}
	\begin{array}{l}
	\displaystyle{	
	{\rm K}(m)=\frac{\pi}{2}\left[1+\frac{m}{4}+\left(\frac{3}{8}\right)^2m^2+O(m^3)\right],
	}
	\\
	\displaystyle{	
	{\rm E} (m)=\frac{\pi}{2}\left[1-\frac{m}{4}-\left(\frac{3}{8}\right)^2\frac{m^2}{3}+O(m^3)\right].
	}
	\end{array}
\end{eqnarray}

In order to prove that $\xi_2$ described by (\ref{sn-syst}) is a monotonically decreasing function, we observe that $K(m)\geq E(m)$, and show that 
\begin{eqnarray}
\label{as_xi}
2\xi_2 K(m)-(\xi_1+\xi_2)E(m)\leq 0, 
\end{eqnarray}
 where $m=(\xi_1-\xi_2)^2/(\xi_1+\xi_2)^2$.
To this end we introduce the elliptic modulus $k=\sqrt{m}$ and rewrite (\ref{as_xi}) as
%\begin{eqnarray}
	%\label{as_xi1}
$	{\cal F}_1(k)\equiv(1-k){\rm K}(k^2)-{\rm E}(k^2)\leq 0$.
%\end{eqnarray}
In order to prove the last formula we use that its left hand side is zero at $k=0$, and its derivative is given by $d{\cal F}_1(k)/dk=-[{\rm E}(k^2)+{\rm K}(k^2)]/(1+k)<0$ (for the last formulas we used~\cite{Lawden}: $d{\rm E}/dk=({\rm E}-{\rm K})/k$ and $d{\rm K}/dk=[{\rm E}(k^2)-k^{\prime 2}{\rm K}(k^2)]/(kk^{\prime 2})$ where $k^{\prime 2}+k^2=1$.

By analogy, in order to show that $\eta_2$ in (\ref{dn-syst}) is monotonically increasing function it is enough to verify that $d{\cal F}_2(k)/dk=-[{\rm K}(k^2)-{\rm E}(k^2)][1-k^\prime]/(kk^\prime)<0$ where $	{\cal F}_2(k)\equiv k^\prime{\rm K}(k^2)-{\rm E}(k^2)\leq 0$.

%%%%%%%%%%%%%%%%%%%%%%%%%

\end{document}